\documentclass[a4paper,12pt]{article}
\usepackage{url} 

\usepackage{amsmath}    
\usepackage{amsfonts}

\usepackage{ifthen}     
\usepackage{bbm}        
\usepackage[pdftex]{hyperref}  

\newcommand{\ie}{{\it i.e.}}
\newcommand{\eg}{{\it e.g.}}

\renewcommand{\d}{\mathrm{d}}
\newcommand{\p}{\partial}
\newcommand{\Order}{\mathrm{O}}

\newcommand{\e}{\mathrm{e}}
\newcommand{\avg}[1]{\left\langle #1 \right\rangle} 
\newcommand{\Tr}{\mathrm{Tr}}

\renewcommand{\b}[1]{{\boldsymbol{#1}}} 

\renewcommand{\r}{{\b{r}}}
\renewcommand{\k}{{\b{k}}}

\newcommand{\K}{{\scriptstyle \b{K}}} 
\newcommand{\nK}{{\scriptstyle K}} 

\newcommand{\0}{\b{0}}
\newcommand{\A}{{\b{A}}}
\newcommand{\y}{{\b{y}}}
\newcommand{\q}{{\b{q}}}

\newcommand{\Wc}{\mathcal{W}^\text{c}}
\newcommand{\Wm}{\mathcal{W}^\text{m}}
\newcommand{\W}{\mathcal{W}}

\newcommand{\equaltime}[4]{\int_{0}^{#1}\!\!\!\!\!\!{\d #2}\!\!\int_{0}^{#3}\!\!\!\!\!\!{\d
        #4}
	\delta(\widetilde{#2}\!-\!\widetilde{#4})\ }
\newcommand{\equaltimeC}[4]{\int_{0}^{#1}\!\!\!\!\!\!{\d #2}\!\!\int_{0}^{#3}\!\!\!\!\!\!{\d
        #4}
    \big(\delta(\widetilde{#2}\!-\!\widetilde{#4})\!\!-\!\!1\big)\ }

\newcommand{\fc}{f^\text{c}} 
\newcommand{\fm}{f^\text{m}} 
\newcommand{\lph}{\lambda_\text{ph}} 
\newcommand{\kB}{k_\text{B}} 

\newcommand{\Y}{\b{Y}}

\newcommand{\D}{\mathrm{D}}

\newcommand{\F}{\mathcal{F}}
\renewcommand{\L}{\mathcal{L}}
\newcommand{\X}{\b{X}}
\newcommand{\Vc}{V^\text{c}}
\newcommand{\Vel}{V^\text{el}}

\newcommand{\FR}{F^\text{R}}

\begin{document}

 \title{\bf Microscopic theory of the Casimir force at thermal equilibrium: large-separation asymptotics}

\author{P. R. Buenzli$^{\ast}$ {\normalsize and}
	Ph. A. Martin$^{\ddag}$\\$^{\ast}$ {\scriptsize Departamento de F\'isica, FCFM,
	Universidad de Chile, Blanco Encalada 2008, Santiago,
	Chile}\\$^{\ddag}${\scriptsize Institute of Theoretical Physics, Swiss Federal
	Institute of Technology Lausanne,}\\{\scriptsize CH-1015 Lausanne EPFL,
	Switzerland}}
\date{}
\maketitle

\abstract{We present an entirely microscopic calculation of the Casimir
force $f(d)$ between two metallic plates in the limit of large separation
$d$. The models of metals consist of mobile quantum charges in thermal
equilibrium with the photon field at positive temperature $T$. Fluctuations
of all degrees of freedom, matter and field, are treated according to the
principles of quantum electrodynamics and statistical physics without
recourse to approximations or intermediate assumptions. Our main result is
the correctness of the asymptotic universal formula $ f(d) \sim
-\frac{\zeta(3) \kB T}{8\pi d^3}$, $d\to\infty$. This supports the fact
that, in the framework of Lifshitz' theory of electromagnetic fluctuations,
transverse electric modes do not contribute in this regime. Moreover the
microscopic origin of universality is seen to rely on perfect screening sum
rules that hold in great generality for conducting media.}

\section{Introduction}
\subsection*{Motivations}
In 1948, Casimir \cite{casimir-plates} predicted that two neutral metallic
plates placed in vacuum at distance $d$ attract one another due to the
electromagnetic field's zero-point fluctuations. In his calculation the
microscopic structure of the conductors is not taken into account. The
latter are treated as macroscopic bodies imposing metallic boundary
conditions on the Maxwell fields.

Early experiments aimed at probing this theoretical prediction remained
inconclusive \cite{sparnaay} until the late 1990s where first experimental
demonstrations were performed \cite{lamoreaux1,lamoreaux2,mohideen-roy},
opening the way to many others (see, \eg,
\cite[Sec. 3.5]{milton-controversies-progress} and
\cite{klimchitskaya-mostepanenko-exp-th} for short reviews). A quantitative
comparison with experiments requires to include a number of effects not
accounted for in Casimir's simple treatment, such as the finite
conductivity of the plates, the roughness of the surfaces and the
dependence on the temperature $T$.

Lifshitz \cite{lifshitz-casimir,landau-lifshitz-EDcontMedia} provided a
first major generalisation by considering plates whose electric properties
are described by a frequency-dependent dielectric function
$\epsilon(\omega)$. The plates are in thermal equilibrium with a stochastic
electromagnetic field, whose random nature is generated by the quantum and
thermal fluctuations of photons and matter. The general force formula
obtained in this way covers in principle a broad diversity of media, and
should be valid for all regimes of temperature T and plate separation
$d$. These regimes are characterized by the single dimensionless parameter
\begin{align}
    \alpha= \frac{\hbar c}{\kB T d} = \frac{\lph}{d}
\end{align}
which measures the ratio of the photon thermal wavelength $\lph\equiv\beta
\hbar c$ to the separation distance $d$ ($\hbar$ is the Planck constant,
$c$ the speed of light, $\kB$ the Boltzmann constant, $\beta=1/\kB T$).

Although the Lifshitz theory (along with its various reformulations
\cite{schwinger2,milonni,bordag-mohideen-mostepanenko-new-developments,milton-controversies-progress,klimchitskaya-mostepanenko-exp-th})
is commonly used to interpret the experimental data, its predictions are
uncertain when applied to \emph{conducting media} at nonzero temperature.
Indeed, the value of the force then depends crucially on the behaviour of
the dielectric function at vanishing frequencies --- a behaviour not
directly accessible to experiments.  This has led to several theoretical
and numerical studies, resulting in a debate that has not yet evolved to a
concensual end (see
\cite{klimchitskaya-mostepanenko-exp-th,lambrecht-neto-reynaud,bezerra-etal,brevik-ellingsen-milton,ellingsen-brevik}
and references cited therein). In short, the controversy amounts to knowing
whether the reflection coefficient of the transverse electric (TE) mode of
the field, $r^\text{TE}(\omega,\k)$ (depending on $\epsilon(\omega)$),
vanishes or not in the limit of zero frequency. In the low-temperature,
small-separation regime $\alpha \gg 1$ one finds the force by unit
surface\footnote{In these formulas, negative/positive terms stand for
attractive/repulsive contributions.}
\begin{align}
    &f(d) \sim
        -\frac{\pi^2 \hbar c}{240 d^4} + \Order(T^4), \phantom{
            \ + \frac{\zeta(3) \kB T}{8\pi\beta d^3}, } \qquad \text{if }
        r^\text{TE}(0,\k)=1, \label{force-lowT1}
        \\
        &f(d) \sim -\frac{\pi^2 \hbar c}{240 d^4}  +
        \frac{\zeta(3) \kB T}{8\pi d^3} + \Order(T^4), \qquad \text{if }
        r^\text{TE}(0,\k)=0, \label{force-lowT2}
\end{align}
while in the high-temperature, large-separation regime $\alpha \ll 1$, one
has:
\begin{align}
&f(d) \sim \displaystyle
        -\frac{\zeta(3) \kB T}{4\pi d^3}, \qquad \text{if } r^\text{TE}(0,\k)=1\label{force-semi-classical1}
        \\& f(d) \sim \displaystyle \!\!\rule{0pt}{4ex}
        -\frac{\zeta(3) \kB T}{8\pi d^3}, \qquad \text{if } r^\text{TE}(0,\k)=0. 
    \label{force-semi-classical2}
\end{align}
In \eqref{force-lowT1}--\eqref{force-lowT2}, the dominant term is in both
cases the standard Casimir result, whereas in the high-temperature,
large-separation regime
\eqref{force-semi-classical1}--\eqref{force-semi-classical2}, one sees a
striking reduction of the force amplitude by a factor $1/2$ when the TE
field modes are assumed not to contribute. Let us add that Formulae
\eqref{force-lowT1}, \eqref{force-semi-classical1} can be obtained under
the assumption of the plasma relation for the low-frequency dielectric
function ($\epsilon(\omega)\sim 1-\omega_\text{p}^2/\omega^2,\;\omega\to
0,\;\omega_\text{p}$ the plasmon frequency), whereas the Drude expression
($\epsilon(\omega)\sim 1-4\pi i \sigma/\omega, \;\sigma$ the conductivity)
leads to the results \eqref{force-lowT2},
\eqref{force-semi-classical2}. The force \eqref{force-semi-classical1} is
also retrieved by extending Casimir's original calculation to finite
temperature between macroscopic plates that are not subject to charge
fluctuation \cite{fierz,mehra}.

Most of the actual debate focused on the finite-temperature corrections in
the low-temperature, small-separation regime. In particular, the fact that
the corrective term linear in $T$ in \eqref{force-lowT2} reflects a nonzero
entropy at $T=0$ consists in an unacceptable violation of the Nernst
postulate for some authors
\cite{klimchitskaya-mostepanenko-exp-th,bezerra-etal}, who thereby favor a
nonvanishing reflection coefficient $r^\text{TE}(0,\k)$. Other authors
\cite{brevik-ellingsen-milton,Hoye-and-al}, however, argue that this linear
correction no longer holds at very low temperature, and favor
\eqref{force-lowT2}, \eqref{force-semi-classical2}.

In this paper, we pronounce on the controversy in the large-separation
regime (with fixed positive temperature). In order to decide which of the
two alternatives \eqref{force-semi-classical1} and
\eqref{force-semi-classical2} is correct, we present a fully microscopic
treatment of the Casimir effect based on the principles of quantum
electrodynamics and statistical mechanics which does not suffer of
intermediate models, assumptions or approximations.\footnote{Experimental
setups in cylinder-plane and parallel plate geometries are currently being
developed with the purposes of discriminating between the different
proposed values for the force
\cite{cylinder-plane-exp-controversy,cylinder-plane-exp-2,plane-plane-exp-proposal}.}
By fully microscopic treatment we mean that all degrees of freedom, matter
and field, are taken into account, contrary to Casimir's original
calculation that ignores particle fluctuations inside the plates.
Microscopic models have been produced to retrieve and justify Lifshitz'
formula in the case of dielectric matter
\cite{renne,hoye-brevik,valeri-sharf}, but conducting media offer more
difficulties as far as one has to deal with screening phenomena due to free
charges and magnetic forces between free currents.

In \cite{buenzli-martin-classical}, we computed the average force by unit
surface between slabs containing purely classical charges and interacting
via the static Coulomb potential, finding
\begin{align}
    f(d) \sim -\frac{\zeta(3) \kB T}{8\pi d^3}, \qquad d\to\infty.
    \label{force8pibeta}
\end{align}
In the letter \cite{buenzli-martin-casimir-letter} we improved the
calculation by considering slabs made of quantum charges and interacting,
in addition to the Coulomb force, with the transverse part of a classical
electromagnetic field. These features do not alter the form
\eqref{force8pibeta} of the large-separation asymptotic force.  Reference
\cite{buenzli-martin-casimir-letter} together with the companion letter
\cite{janco-samaj-letter} stress the importance of including in the
calculation the effects of the charge fluctuations in the metals, which are
responsible for reducing the asymptotic force amplitude
\eqref{force-semi-classical1} by the factor $1/2$.  It is also understood
why the entirely classical model treated in \cite{buenzli-martin-classical}
correctly predicts the high-temperature, large-separation result
\eqref{force8pibeta}: this is a consequence of the Bohr--van Leeuwen
theorem. The theorem states that in classical systems at thermal
equilibrium, matter decouples from the transverse electromagnetic
field. Since high-temperature conducting phases tend to behave classically,
the corresponding Casimir force will be determined at leading order by
purely electrostatic interactions.

\subsection*{Statement of results} 
The present paper addresses the question of the Casimir force in the
general framework of nonrelativistic thermal quantum electrodynamics
(TQED), namely nonrelativistic quantum charges in interaction with the
quantized electromagnetic field. The model (described in more detail in
Section 2) consists of mobile quantum charges confined in two slabs $A$ and
$B$ of thickness $a$ and $b$ with lateral faces of surface $L^2$, set at
distance $d$ from each other. The charges interact with a quantum
electromagnetic field enclosed in a large box $\Lambda$. The Hamiltonian
$H_{\Lambda,L,d}$ of the system is specified in Formula (\ref{hamiltonian})
of Section 2. The photons and the particles are supposed to be in thermal
equilibrium at temperature $T$, so that all the relevant information is
contained in the grand-canonical potential $ \Phi_{\Lambda,L,d}$ associated
with the Hamiltonian $H_{\Lambda,L,d}$. The average force by unit surface
exerted between the plates is defined by the rate of change occasioned in
$\Phi_{\Lambda,L,d}$ when varying the separating distance $d$:
\begin{align}
    f_{\Lambda,L}(d) = - \frac{1}{L^2} \frac{\p}{\p d}\Phi_{\Lambda,L,d}.
\end{align}
The Casimir force is defined as
\begin{align}
    f(d) \equiv \lim_{L^2\to\mathbb{R}^2}\lim_{\Lambda\to\mathbb{R}^3}
    f_{\Lambda,L}(d), \label{casimir-force-def}
\end{align}
where the thermodynamic limit of the system is taken in two stages. We
first let the box enclosing the field $\Lambda\to\mathbb{R}^3$, and then
extend the plates' surfaces $L^2 \to \mathbb{R}^2$. The plates' thicknesses
$a$ and $b$ are kept finite.  The main result is that at any fixed
temperature $T>0$ (such that the thermal energy $k_BT$ is much less than
the rest mass energies $mc^2$ of the particles), the large-separation
asymptotic force is again given by Formula \eqref{force8pibeta}. The
amplitude is linear in $T$, independent of the Planck constant $\hbar$ and
of the speed of light $c$, and universal with respect to the microscopic
constitution of the plates.  Nonuniversal contributions and contributions
depending on $\hbar$ and $c$, will only occur at the next order
$\Order(d^{-4})$ in the large-separation expansion of the force. This
result not only validates from first principles the second alternative in
Eqs. \eqref{force-semi-classical1}, \eqref{force-semi-classical2}
(associated with the vanishing of the reflection coefficient for TE modes),
but also establishes universality of the leading term \eqref{force8pibeta}
on a microscopic basis.  Since our methods might be not quite familiar, we
summarize the main steps of our derivation.

\subsection*{Methods}
\subsubsection*{Casting the quantum system in a classical-like form: the space of loops}
When we have a conducting medium it is of utmost importance to deal
properly with the collective screening effects. The idea is to cast the
quantum system in a form which is as close as possible to that of a
classical system of charges, where we have for instance the well-developed
Debye-H\"uckel theory of screening. To this effect we introduce a joint
functional integral representation of the Gibbs weight associated with the
total Hamiltonian of matter and field. In this formalism, developed in
\cite{buenzli-martin-ryser} and recalled in Section 2, quantum particles
appear as extended objects consisting of random closed wires $\L$ (called
loops) carrying both a charge and a current. The loop size, which is
measured by the thermal de Broglie wavelength, reflects the intrinsic
quantum fluctuation of the particle. There are two kinds of pairwise
interactions between loops. The first one $\Vc(\L_ i,\L_ j)$ originates
from the electrostatic (Coulomb) potential between charges, Formula
\eqref{Vc}. The second one, $\Wm(\L_ i,\L_ j)$ (Formula \eqref{Wm}) is
called the magnetic potential.  It is an effective interaction resulting
from integrating out the field degrees of freedom: one can figure it as
current interactions between the loops mediated by the transverse part of
the electromagnetic field. At this point, although being an exact
representation of the quantum TQED system, the statistical mechanics of
loops has a classical-like structure which enables a convenient application
of the methods of classical statistical mechanics.
                                                                                                                                                                                                                                                                                                                                                                                                                                                                                                                                                                                                                                                                                                     
\subsubsection*{Expressing the Casimir force in terms of loops}
The force between two loops is given, as in classical physics, by the
gradients of the potentials $\partial_x\Vc(\L_ i,\L_ j)$ and
$\partial_x\Wm(\L_ i,\L_ j)$ (along the $x$-axis perpendicular to the
plates). The average force, as usual, is obtained by averaging these forces
with the equilibrium correlation function $\rho^{(2)}(\L_ i,\L_ j)$ between
two loops. The precise expression is found in Formula
(\ref{force-ursell-cap}) in Section 3 where we have also singled out the
proper Casimir force due to fluctuations. The additional part, called here
capacitor force, is the direct Coulomb force that would occur between
globally nonneutral plates. One can benefit from the translational
invariance along the ${\bf y}$ directions parallel to the plates by using
the correponding two-dimensional Fourier variable ${\bf k}$ and scale it as
${\bf k}={\bf q}/d$ where ${\bf q}$ is now a dimensionless Fourier
variable. The scaling trivially provides a prefactor $1/d^2$ in the force
Formula (\ref{force-ursell-loop}). The remaining $d$-dependence remains
embedded in the microscopic expressions of the forces and correlations
between loops.

\subsubsection*{Screening of the electrostatic interactions}
This is the subject of Section 4. The main observation is that the Coulomb
interaction between loops can be decomposed into $\Vc=\Vel+\Wc$ (see
Formulas (\ref{Vel}) and (\ref{Wc})). Here $\Vel$ is the genuine classical
electrostatic interaction between charged wires whereas $\Wc$ incorporates
the proper effect of the quantum nature of the particles manifested by the
fluctuations of the loops. One can easily extend to $\Vel $ the standard
ideas of the classical Debye-H\"uckel theory, providing an effective
resummed potential $\Phi$ that becomes integrable at large distances (in
the planar geometry one is concerned by the integrability in the ${\bf y}$
directions along the plates, see Appendix C). One is therefore left with
the screened potential $\Phi$ together with the additional interactions
$\Wc$ and $\Wm$. The latter would not be present for classical charges:
they result from the intrinsic quantum fluctuations of the particles and
behave as electric and magnetic dipole interaction at large distance. At
this point on can use in the space of loops the methods of Mayer expansion
and integral equations well developed in the context of classical Coulomb
fluids. Of particular importance is the perfect screening sum rule stating
that any specified loop is surrounded by a screening cloud of loops whose
total charge compensates that of the specified loop. This imposes an exact
integral constraint on the two-loop correlation function, Formula
(\ref{loop-perfect-screening}), that turns out to be at the origin of the
universality of the Casimir force.

\subsubsection*{The large-separation asymptotics}
Apart from the obvious $1/d^2$ scaling factor, the $d$-dependence of the
force has to be extracted from the forces and correlations between
loops. The Coulomb part of the force (expressed in terms of the ${\bf q}$
transverse Fourier variable) has a non vanishing limit as $d\to\infty$
whereas the magnetic force vanishes as $1/d^2$ and the correlation between
the plates as $1/d$ (Section 6). This implies that the dominant term decays
as $1/d^3$ and does not involve a direct contribution of the magnetic
force. To obtain the amplitude of this $1/d^3$ term it is necessary to
determine the exact asymptotic form of the correlation. The latter is seen
to be made of two terms (Formulae \eqref{hAB-asymptotic-FAB} and
\eqref{hAB-asymptotic-WAB}) decaying as $1/d$ times a product of certain
arrangements of correlations pertaining to the individual plates.
When this is introduced in the force Formula (\ref{force-ursell-loop}) one
discovers that the perfect screening sum rules in each of the plates wash
out all details of the microscopic structure of the conductors, thereby
leading to the wonderfully-simple result \eqref{force8pibeta} and providing
a physical explanation of universality (Section 5). The analysis of the
correlation in Subsections 6.1 and 6.3, if somewhat lengthy, uses common
reasoning in terms of Mayer graph diagramatics. It basically reveals that
the electric and magnetic dipolar potentials between the loops $\Wc$ and
$\Wm$ do not eventually contribute to the dominant $1/d^3$ term of the
force. This term is entirely due to electrostatics and screening, so
explaining why the purely classical model of
\cite{buenzli-martin-classical} gives the correct result. The complexity of
the full quantum mechanical treatment presented in this paper contrasts
with the simplicity of the result, a fact that we could not foresee right
away. More comments and perspectives are offered in the Concluding remarks
in Section 7.

\section{Description of the system}
We consider two parallel slabs $A=[-a,0]\times L^2$ and $B_d=[d,b+d]\times
L^2$ with thickness $a$ and $b$ and lateral surface $L^2$. The $x$-axis is
perpendicular to the plates, the inner face of slab $A$ being fixed at
$x=0$ while the slab $B_d$ is set at a distance $d$ from it. The slabs
contain nonrelativistic point particles of several species $\gamma$
(electrons, ions, nuclei) with charges $e_{\gamma}$, masses $m_\gamma$,
spins $s_\gamma$ and appropriate statistics.  These particles are confined
by walls without electrical properties in the two separate regions and no
exchange is possible from one slab to the other. Particles in one plate are
always distinguishable from alike particles in the other plate. To ensure
the global neutrality of each plate, we impose
\begin{align}
    \sum_a e_{\gamma_a} = \sum_b e_{\gamma_b} = 0, \label{neutrality}
\end{align}
where the sums are carried over particles in $A$ and $B_d$, respectively.

This system of interacting charges is coupled to a quantum electromagnetic
field which is itself enclosed into a larger box $\Lambda$ englobing both
plates. The $N$-particle Hamiltonian reads
\begin{align}
    H_{\Lambda,L,d} = &\sum_{i=1}^N \!\frac{1}{2m_{\gamma_i}}\Big(\b{P}_i \!-\!
    \frac{e_{\gamma_i}}{c} \A(\r_i) \!\Big)^2 + \sum_{i<j} e_{\gamma_i}
        e_{\gamma_j} v(\r_i-\r_j) \notag
        \\&+ \sum_{i=1}^N
    V^\text{walls}(\r_i, \gamma_i) + H_{0,\Lambda}^\text{rad} \label{hamiltonian}
\end{align}
with $v(\r_i-\r_j)$ the static Coulomb potential
\begin{align}
	v(\r_i-\r_j)=\frac{1}{|\r_i-\r_j|}. \label{v}
\end{align}
As is common in atomic physics when matter is nonrelativistic and
high-energy processes are neglected, we use the Coulomb gauge and
electrostatic Gaussian units \cite{cohen-photons-atoms}. The Coulomb gauge
has the advantage of clearly disentangling electrostatic and magnetic
couplings in the Hamiltonian. The divergence-free vector potential $\A(\r)$
is supposed to satisfy periodic boundary conditions on the sides of the box
$\Lambda$. Its expansion in Fourier modes $\K$ is given by
\begin{align}
    &\A(\r) = \left(\frac{4\pi\hbar c^2}{\Lambda}\right)^{1/2} \sum_{\K,\lambda}
    g(\K) \frac{\b{e}_{\K,\lambda}}{\sqrt{2\omega_\K}}
    \big( a^\ast_{\K,\lambda} \e^{-i\K\cdot \r} + a_{\K,\lambda}
    \e^{i\K\cdot\r} \big), \label{vectorpotential}
\end{align}
where $a^\ast_{\K,\lambda}, a_{\K,\lambda}$ are the creation and
annihilation operators for the mode $\K,\lambda$ of frequency
$\omega_\K=c|\K|$ with commutation relations $[a_{\K,\lambda},
a^\ast_{\K',\lambda'}]=\delta_{\K,\K'}\delta_{\lambda,\lambda'}$;
$\b{e}_{\K,\lambda}, \lambda=1,2$ are the polarization vectors; $g(\K)$,
$g(\b{0})=1$, is a real, spherically-symmetric, and smooth form factor
taking care of ultraviolet divergencies. It is supposed to decay rapidly to
$0$ beyond the characteristic wavenumber $\nK_\text{cut} \equiv
\frac{2\pi}{\lambda_\text{cut}}=\frac{\bar{m}}{\hbar c}$ where $\bar{m}$ is
an average particle mass. The term $H_{0,\Lambda}^\text{rad}$ in
\eqref{hamiltonian} is the free field Hamiltonian
\begin{align}
    &H_{0,\Lambda}^\text{rad} = \sum_{\K,\lambda} \hbar \omega_\K\ 
    a^\ast_{\K,\lambda}a_{\K,\lambda}. \label{freefieldhamiltonian}
\end{align}
The wall potential $V^\text{walls}(\r_i,\gamma_i)$ confines the particles
either to slab $A$ or to slab $B_d$, depending on whether $\gamma_i$
designates a species in $A$ or $B_d$. Note that we neglect spin--field
couplings in this model (see comments in the Concluding remarks).

The states of this system of particles and field are supposed to be
thermalised at the inverse temperature $\beta=(\kB T)^{-1}$, and statistical
averages, denoted by $\avg{\ldots}$, are taken with the usual Gibbs weight
$\e^{-\beta H_{\Lambda,L}}$. We introduce the finite-volume grand-canonical
potential of the full system
\begin{align}
    \Phi_{\Lambda,L,d} = -\kB T \ln \Tr\ \e^{-\beta (H_{\Lambda,L}-
    \b{\mu} \cdot\b{N}})
    \label{grand-potential}
\end{align}
where the trace $\Tr = \Tr_\text{mat} \Tr_\text{rad}$ is carried over
particles' and field's degrees of freedom. Here
$\b{\mu}=\{\mu_{\gamma_a},\mu_{\gamma_b}\}$ is the collection of chemical
potentials that fix the average particle densities in each of the plates
and $\b{N}=\{N_{\gamma_a},N_{\gamma_b}\}$ are the corresponding particle
numbers. In (\ref{grand-potential}) $\Tr_\text{mat}$ is carried out only on
neutral configurations in each plate.
The average force by unit surface exerted between infinitely extended plates immersed in the electromagnetic field is then defined by Formula (\ref{casimir-force-def}), the temperature and
chemical potentials being fixed.

In addition to the separation $d$, there is a number of other
characteristic lengths in the system, in particular, the thermal wavelength
of photon $\lambda_\text{ph}=\beta\hbar c$ and of particles $\lambda_\text{mat}=\hbar
\sqrt{\beta/\bar{m}}$. Moreover, the plates are assumed to be conducting,
and therefore characterized by a screening length
$\lambda_\text{screen}$. Our derivation holds for the following hierarchy
of lengths:
\begin{gather}
\lambda_\text{cut}=\frac{\lambda_\text{mat}}{\sqrt{\beta m c^2}} \ll
\lambda_\text{mat} \ll \lambda_\text{ph} = \sqrt{\beta m c^2}
\lambda_\text{mat} \ll d \label{ineq1},
\\ \lambda_\text{screen} \ll a,b \ll d. \label{ineq2}
\end{gather}
The first set of inequalities is necessary for the consistency of the
nonrelativistic treatment of matter, which requires $\beta \bar{m} c^2 \gg
1$ (the thermal energy is much smaller than the rest mass energy of the
particles). Inequality \eqref{ineq2} means that the plates' thickness
should be large enough for allowing the screening mechanisms to take place
inside the conductors. Finally, the conductors will be assumed to be
invariant under translations and rotations in the plate directions.

\subsubsection*{Loop formalism}
Our analysis relies on the formalism developed in
\cite{buenzli-martin-ryser}, based on a joint functional representation of
both matter and field. In this formalism, the field degrees of freedom can
be integrated out exactly. Then, the particle variables live in an
auxiliary classical-like phase space whose elements are loops of random
shape (for the statistical mechanics of charged loops, see the review \cite{brydges-martin}, Chap. V and references therein). A loop $\L=(\r,\chi)$ is specified by a position $\r$ in space and a
number of internal degrees of freedom $\chi=(\gamma,p,\X(\cdot))$
consisting of a species index $\gamma$, a charge number $p \in 1,2,3,...$
and a closed Brownian path $s\mapsto \X(s)$, $s\in [0,p]$,
$\X(0)=\X(p)$. The loop's shape $\X(s)$ is a Gaussian stochastic process
(Brownian bridge) whose functional integral has unit normalization, zero
mean, and covariance given by
\begin{align}
	\int\!\!\D(\X) X^\mu(s)X^\nu(s')= \delta_{\mu\nu}\ p
	\left(\text{min}\left\{\frac{s}{p},\frac{s'}{p'}\right\} - \frac{s}{p}
	\frac{s'}{p'} \right). \label{X-covariance}
\end{align}
The loop's path is
\begin{align}
	\r^{[s]} \equiv \r+\lambda_\gamma \X(s), \quad 0\leq s\leq p
	\label{loop-path}
\end{align}
where $\lambda_\gamma = \hbar \sqrt{\beta/m_\gamma}$ is the de Broglie thermal
wavelength. The occurrence of the Brownian path results from the
Feynman--Kac path integral and $\lambda_\gamma$ gives the extension of the
quantum particle's fluctuation. The number $p$ accounts for the quantum
statistics of the species $\gamma$. It corresponds to grouping together $p$
particles that are permuted accordingly to a cyclic permutation of length
$p$.

The pairwise interaction $e_{\gamma_i} e_{\gamma_j} V(\b i, \b j)$ between
two loops $\b i\equiv\L_i$ and $\b j\equiv\L_j$ is the sum of two
contributions
\begin{align}
    e_{\gamma_i} e_{\gamma_j} V(\b i, \b j) = e_{\gamma_i}e_{\gamma_j}
    \big[\Vc(\b i,\b j) + \Wm(\b i,\b j)\big].
	\label{V}
\end{align}
The first contribution, inherited from the Coulomb potential, is
\begin{align}
	&\Vc(\b i,\b j) = \equaltime{p_i}{s_i}{p_j}{s_j}
    \frac{1}{\big|\r_i^{[s_i]}-\r_j^{[s_j]}\big|}, \label{Vc}
\end{align}
where $\tilde s= s\ \text{mod}\ 1$ and
$\delta(\widetilde{s_i}-\widetilde{s_j})$ takes into account the equal-time
constraint imposed by the Feynman--Kac formula. The second contribution is
the effective potential resulting from the elimination of the field's
degrees of freedom. We call it the magnetic potential. It is given in
Fourier representation by Formula (66)-(67) of \cite{buenzli-martin-ryser}:
\begin{align}
    \Wm(\b i,\b j) = &\int\!\!\! \frac{\d\K }{(2\pi)^{3}}\, \e^{i
    \K\cdot(\r_i-\r_j)}\ \Wm(\chi_i,\chi_j,\K), \label{Wm}
	\\\Wm(\chi_i,\chi_ j,\K)=&
 \frac{1}{\beta
        \sqrt{m_{\gamma_i}m_{\gamma_j}}c^{2}}\int_{0}^{p_{i}}\!\!\! \d
    X_{i}^\mu(s_i)\, \e^{i\K\cdot\lambda_{\gamma_{i}}\X_{i}(s_i)}
       \notag
    \\\times \int_{0}^{p_{j}}\!\!\! \d &X_{j}^\nu(s_j)\, \e^{-i\K\cdot\lambda_{\gamma_{j}}\X_{j}(s_j)}
   \frac{4\pi g^2(\nK)}{\nK^2}\delta_{\mu\nu}^{\text
   tr}(\K)\mathcal{Q}(\nK,\tilde{s_i}\!-\!\tilde{s_j}),\notag
\end{align}
with
\begin{align}
&\mathcal{Q}(\nK,\tilde{s_i}\!-\!\tilde{s_j}) \equiv
\frac{\lambda_{\text{ph}}\nK}{2\sinh(
\lambda_{\text{ph}}\nK/2)}\cosh[\lambda_{\text{ph}}\nK(|\tilde{s_i}\!-\!\tilde{s_j}|-1/2)]\label{Q},
\end{align}
and
\begin{align}
& \delta_{\mu\nu}^\text{tr}(\K) \equiv \delta_{\mu\nu} - \frac{k_\mu
		k_\nu}{\nK^2} \qquad (\K=\{k_\mu\}_{\mu=1}^3, \ \nK=|\K|) \label{delta-transverse}
\end{align}
is the transverse Kronecker function. In \eqref{Wm}, $\int_0^p \d X^\mu(s)$
are stochastic line integrals along the loop shape. The function
$\mathcal{Q}$, depending only on $\lambda_\text{ph}$, is the manifestation
of the quantum photon field. This formula holds when the field region
$\Lambda$ has been extended to infinity, replacing the discrete sum on
Fourier modes by an integral.

Written in terms of loop variables, the grand-canonical partition function
of the full system, normalised by that of the free radiation field, has a
classical structure:
\begin{align}
	 &\Xi_{L,d} = \lim_{\Lambda\to\mathbb{R}^3} \frac{\Tr\ \e^{-\beta
	( H_{\Lambda,L}-\b{\mu} \cdot\b{N})}}{\Tr\ \e^{-\beta H_{0,\Lambda}^\text{rad}}} \notag \\&=
	 \!\!\sum_{n_A=0}^\infty\!\! \frac{1}{n_A!}  \!\sum_{n_B=0}^\infty\!\!
	 \frac{1}{n_B!}  \!\int_{\!A} {\prod_{a}^{n_A}\d\L_a\, z(\L_a)}
	 \!\int_{\!B_d} {\prod_b^{n_B}\d\L_b\, z(\L_b)} \, \e^{-\beta
	 U(\{\L_a\}, \{\L_b\})}.
\label{loop-partition-function}
\end{align}
In \eqref{loop-partition-function}, the loop integration $\int_A\! \d \L_a
= \int_A\! \d \r_a\, \d \chi_a = \int_A\! \d\r_a \sum_{\gamma_a}\! \sum_{p_a}
\!\!\D(\X_a)$ is carried over paths $\r_a^{[s]}$ entirely contained
in slab $A$, respectively, $\int_{B_d}\! \d \L_b$ over paths in slab
$B_d$. This corresponds to choosing hard walls on the faces of the slabs,
\ie, Dirichlet boundary conditions for the particle wavefunctions. 
In (\ref{loop-partition-function}) the sums run only on neutral configurations of loops in each slab.

The
total loop energy $U$ can be separated into intra and interplate
contributions:
\begin{align}
	U = U_A + U_{B_d} + U_{AB_d}
\label{pot.energy}
\end{align}
where
\begin{align}
	U_A = \sum_{\b i, \b j \in A}  e_{\gamma_i} e_{\gamma_j}V(\b i, \b j)
\end{align}
is the sum of interactions occurring among loops confined into slab $A$
(likewise for $U_{B_d}$ in slab $B_d$), and
\begin{align}
	U_{AB_d} = \sum_{\b i \in A} \sum_{\b j \in B_d} e_{\gamma_i} e_{\gamma_j} V(\b i, \b j)
\end{align}
is the interaction energy between the two plates. Moreover, each loop is
equipped with an effective activity $z(\L)$ containing the loop self-energy
$e_\gamma^2 V(\L,\L)$:
\begin{align}
    z(\L) = \frac{(2 s_\gamma + 1) (\eta_{\gamma})^{p-1}}{p} \frac{(\e^{\beta
            \mu_\gamma})^p}{(2\pi p \lambda_{\gamma}^2)^{3/2}} \e^{-\beta
        \frac{e_\gamma^2}{2} V(\L,\L)}.\label{loop-activity}
\end{align}
The factor $(2 s_\gamma +1)$ accounts for the spin-degeneracy of the energy
levels, $\mu_\gamma$ is the chemical potential of species $\gamma$ and
$\eta_\gamma = \pm 1$ for bosonic/fermionic species.

We stress that although the loop partition function
\eqref{loop-partition-function} has a classical form, it is a
mathematically exact representation of the original grand-canonical
partition function of the system of quantum charges and photons as defined
by the Hamiltonian \eqref{hamiltonian}.

\section{The Casimir force}
In view of \eqref{grand-potential}-\eqref{casimir-force-def} and the fact
that the free-field partition function does not depend on $d$, the Casimir
force expressed with the help of the loop partition function
\eqref{loop-partition-function} reads
\begin{align}
    f(d) &= \lim_{L\to\infty}\lim_{\Lambda\to\mathbb{R}^3} \frac{\kB
    T}{L^2} \frac{\p}{\p d} \left[ \ln \Tr\ \e^{-\beta (H_{\Lambda,L}-\b{\mu} \cdot\b{N})} - \ln
    \Tr\ \e^{-\beta H_{0,\Lambda}^\text{rad}} \right] \notag
    \\ &= \lim_{L\to\infty} \frac{\kB T}{L^2}\ \frac{\tfrac{\p}{\p d}
    \Xi_{L,d}}{\Xi_{L,d}}. \label{force-field-qm}
\end{align}
The dependence upon $d$ in the partition function
\eqref{loop-partition-function} occurs only  in the confinement of
the loops in the slab $B_d$. For the rest of the paper, it is convenient to shift the positional
integration variable $x_b \in [d,b+d]$ of a loop in $B_d$ to $x_b-d$, so
that slab $B_d$ is moved to the fixed region $B=[0,b]\times L^2$. Then,
the $d$-dependence is transfered to the interaction potential between
slabs:
\begin{align}
	V(\L_a,\L_b),\ \ x_b \in [d,b+d]\quad \mapsto\quad V(\L_a,\L_b + d),\ \
	x_b \in [0,b],
\label{shift}
\end{align}
where $\L_b + d$ is the loop $\L_b$ shifted along the $x$-axis from $x_b$ to
$x_b +d$. This amounts to measure the positions in slab $B_d$ from its
inner face.  To abbreviate the notation, we set
\begin{align}
	V_{AB}(\L_a, \L_b) \equiv V(\L_a, \L_b+d) \label{AB-notation}.
\end{align}
From now on it will be implicitly understood that $ V_{AB}(\L_a, \L_b)$
depends on $d$ according to \eqref{AB-notation} and that in forthcoming
integrals the path of the loop $\L_a$ is restricted to the fixed slab $A$
of volume $[-a,0]\times L^2$ while that of $\L_b$ to the fixed slab B of
volume $[0,b]\times L^2$.
 
Differentiating with respect to $d$ in \eqref{force-field-qm} is equivalent
to differentiating the potential $V(\L_a, \L_b +d)$ with respect to $x_b$,
or with respect to $-x_a$ (since the dependence on the $x$ components is
$x_a-x_b-d$). This brings in the average force along $x$
\begin{align}
    f(d) &= \lim_{L\to\infty} \frac{1}{L^2}\avg{\sum_a^A \sum_b^B
        e_{\gamma_a}e_{\gamma_b}(\p_{x_a} \Vc_{AB} + \p_{x_a}
        \Wm_{AB})(\L_a,\L_b)}_\text{loops}.
	\label{force-average-loops}
\end{align}
The bracket $\avg{\cdots}_\text{loops}$ denotes the grand-canonical
statistical average in the phase space of loops with activities
\eqref{loop-activity} and with respect to the Gibbs weight $\e^{-\beta U}$
associated to the loop potential energy \eqref{pot.energy}.  Since the
force is a two-body observable, its thermal average can be expressed as
an integral over the two-loop correlation $\rho^{(2)}_{L}$ between a loop
in $A$ and a loop in $B$:\footnote{Loop correlation functions are defined
similarly to particle density correlation functions, see
\cite{brydges-martin}, Chap. V.}
\begin{align}
        f(d) = \lim_{L\to\infty} \frac{1}{L^2} \!\!\int_{\! A}\!\! {\d \b
        1}\!\! \int_{\! B}\!\!\! {\d \b 2}\, e_{\gamma_1}
        e_{\gamma_2} (\p_{x_1} \Vc_{AB} + \p_{x_1} \Wm_{AB})(\b 1, \b 2) \,
        \rho_{AB,L}^{(2)}(\b 1, \b 2),
    \label{loop-micro-force-c-m}
\end{align}
where we have followed the notation \eqref{AB-notation},
\begin{align}
	\rho_{AB,L}^{(2)}(\L_1,\L_2) \equiv \rho_L^{(2)}(\L_1, \L_2+d),
	\label{AB-notation2}
\end{align}
and the $\d\b 1$ integration is carried on loops in $A$, respectively, the $\d\b
2$ integration on loops in $B$.

At this stage we take the limit $L\to\infty$ of infinite plate
surfaces. Since the conductors are assumed to become homogeneous in the
$\y=(y,z)$ plane of the plates, the two-loop correlation function tends to
a function $\rho_{AB}^{(2)}(1,2,|\y_1-\y_2|)$. In this geometry, it is
convenient to decompose $\r_1=(x_1,\y_1)$, and
\begin{align}
	\b 1 = (1,\y_1),\quad \d \b 1=\d 1\ \d\y_1,\quad \d 1= \d x_1 \d\chi_1
 \label{loop-notation}
\end{align} 
where $1=(x_1,\chi_1)$ denotes the position along $x$ of the loop $\L_1$
and its internal degrees of freedom (likewise for $2$).  In the limit, the
factor $1/L^2$ cancels with one of the $\y$-integral in
\eqref{loop-micro-force-c-m}, yielding
\begin{align}
    f(d) &= \!\!\int_{\! A}\!\!\! {\d 1}\!\! \int_{\! B} \!\!\! {\d 2}\!\!
    \int\!\!\! {\d\y}\  e_{\gamma_1} e_{\gamma_2} (\p_{x_1} \Vc_{AB}
        + \p_{x_1} \Wm_{AB})(1, 2, \y) \rho_{AB}^{(2)}(1, 2,
        \y) \label{force-loops-y}.
\end{align}
We introduce in \eqref{force-loops-y} the loop Ursell function $h(\b 1, \b
2)$ defined in the usual way by
\begin{align}
    \rho(\b 1)\rho(\b 2) h(\b 1, \b 2) \equiv \rho^{(2)}(\b 1, \b
    2) - \rho(\b 1)\rho(\b 2), \label{ursell-loop}
\end{align}
with $\rho(\L)$ the loop density, so that
\begin{align}
    f(d) = &\int_{\! A}\!\!\!{\d 1}\!\!
    \int_{\! B}\!\!\!{\d 2}\!\!  \int\!\!\!{\d\y}\ e_{\gamma_1}
    e_{\gamma_2} (\p_{x_1} \Vc_{AB}
        + \p_{x_1} \Wm_{AB})(1, 2, \y) \notag
    \\&\times\rho_A(1)\rho_B(2) h_{AB}(1, 2, \y) + f_\text{cap}(d),
\label{force-ursell-cap}
\end{align}
where again $h_{AB}(\L_1,\L_2 )=h(\L_1,\L_2+d)$ and $\rho_A(\L_1)=\rho(\L_1),
\rho_B(\L_2)=\rho(\L_2+d)$ according to the notation \eqref{AB-notation},
\eqref{AB-notation2}.  The capacitor force $f_\text{cap}(d)= \fc_\text{cap}
+ \fm_\text{cap}$ comes from the subtracted product of the plates' density
in \eqref{ursell-loop}.  We show in Appendix \ref{appx:capacitor-force}
that the electrostatic part $\fc_\text{cap}(d)$ due to the term
$\p_{x_1}\Vc_{AB}$ reduces to
\begin{align}
	\fc_\text{cap}(d) = 2\pi \left[\int_{-a}^0\!\!\!{\d x_1}\, c_A(x_1) \right]
        \left[\int_0^b\!\!\!{\d x_2}\, c_B(x_2) \right],
        \label{capacitor-force}
\end{align}
where $c_A(x_1)$ and $c_B(x_2)$ are the mean charge densities in plate $A$
and $B$. It corresponds to the standard force (in Gaussian units) between a
capacitor's plates whose surface charge densities are respectively
$\int_{-a}^0\!  \d x_1 c_A(x_1)$ and $\int_0^b\! \d x_2 c_B(x_2)$.  In this
work, we assume strict neutrality in virtue of \eqref{neutrality}, so that
$\fc_\text{cap}(d)\equiv 0$.  Moreover, we also show in Appendix
\ref{appx:capacitor-force} that the magnetic contribution
$\fm_\text{cap}(d)$ decays faster than any inverse power of $d$. In the
sequel, we will thus drop the capacitor force and focus only on the first
term of \eqref{force-ursell-cap}, which is the proper Casimir force
generated by fluctuations.

There is a noteworthy simplification in the electrostatic part due to
$\p_{x_1} \Vc_{AB}(1,2,\y)$ in the Casimir force. Namely, one can omit all
multipolar contributions in the loop Coulomb force $\p_{x_1} e_{\gamma_1}
e_{\gamma_2} \Vc(\b 1, \b 2)$ (see \eqref{Vc}) replacing it by its pure
monopole term $\p_{x_1} p_1 p_2 e_{\gamma_1}e_{\gamma_2}
\frac{1}{|\r_1-\r_2|}=  p_1 p_2 e_{\gamma_1}e_{\gamma_2}
\p_{x_1} v(\r_1-\r_2)$. The underlying reason is that the electric force expressed
in terms of the original two-point particle correlation function involves
the average standard Coulomb force $\p_{x_1} v(\r_1-\r_2)$ between point
charges. The equivalence with the present formulation in terms of loops is
given in Appendix \ref{appx:electro-force}.

Finally, we represent the remaining $\y$-integral in the two-dimensional
(transverse) Fourier space $\k$ and introduce the dimensionless variable
$\q=\k d$:
\begin{align}
    f(d) = & \frac{1}{d^2} \!\!\int_{\! A}\!\!\!{\d 1}\!\!
    \int_{\! B}\!\!\!{\d 2}\!\!  \int\!\!\!{\frac{\d\q}{(2\pi)^2}} e_{\gamma_1}
    e_{\gamma_2} \left(p_1 p_2\, \p_{x_1} v_{AB} + \p_{x_1}
    \Wm_{AB}\right)(1, 2, \tfrac{\q}{d}) \notag
    \\&\times\rho_A(1)\rho_B(2) h_{AB}(1, 2, \tfrac{\q}{d})  
	, \label{force-ursell-loop}
\end{align}
where
\begin{align}
	\p_{x_1} v_{AB}(1,2,\tfrac{\q}{d}) = \p_{x_1} \!\!\int\!\! \d\y\  \frac{\e^{i\frac{\q}{d}\cdot \y}}{\sqrt{(x_1-x_2-d)^2+\y^2}} =
	2\pi\,\e^{-q}\,\e^{-{q(x_2-x_1)}/{d}}. \label{coulomb-force}
\end{align}
The general formula \eqref{force-ursell-loop} is an expression structurally
similar to the one developed in the purely classical model
\cite[Form. (29)]{buenzli-martin-classical}. It reduces to it when charges
are classical and the field is switched off. The main purpose is now to
extract the $d$-dependence of the Ursell function
$h_{AB}(1,2,\tfrac{\q}{d})$, which embodies all correlations between the
two plates.

\section{Screening of the electrostatic interaction}
The Ursell function can be conveniently analysed by performing a Mayer
expansion in the phase space of loops. The Mayer bonds $f(\b i, \b j) =
\e^{-\beta e_{\gamma_i} e_{\gamma_j} V(\b i, \b j)}-1$ are built from the
basic loop-loop interaction \eqref{V}. The Coulombic part of \eqref{V}
decays as $r^{-1}$ so that the Mayer bond is not integrable. In order to
remedy to this nonintegrability, it is necessary to take screening effects
into account. To this end, we first make the following observation. From
the Feynman--Kac formula the potential \eqref{Vc} inherits the
quantum-mechanical equal-time constraint: \ie, every element of charge
$e_{\gamma_i} \lambda_{\gamma_i} \d \X_i(s_i)$ of the first loop does not
interact with every other element $e_{\gamma_j} \lambda_{\gamma_j} \d
\X_j(s_j)$ as would be the case in classical physics, but the interaction
takes place only if $s_1=s_2$. It is therefore of interest to split $\Vc$
into $\Vel+\Wc$, where
\begin{gather}
      \Vel(\b i,\b j) = \int_{0}^{p_i}\!\!\!\!\!\!{\d s_i}\!\!
     \int_{0}^{p_j}\!\!\!\!\!\!{\d s_j}
     \frac{1}{\big| \r_i^{[s_i]}-\r_j^{[s_j]}\big |}, \label{Vel} 
    \\ \Wc(\b i,\b j) = \equaltimeC{p_i}{s_i}{p_j}{s_j}
    \frac{1}{\big|\r_i^{[s_i]}-\r_j^{[s_j]}\big|}. \label{Wc}
\end{gather}
The contribution $\Vel$ is the genuine classical Coulomb interaction
between two uniformly charged wires of shapes $\r_i^{[s_i]}$ and
$\r_j^{[s_j]}$, whereas the quantum-mechanical constraint appears in $\Wc$.

Now, the complete two-loop potential \eqref{V} reads
\begin{align}
	V=\Vel + \W, \quad \text{with } \W=\Wc+\Wm \label{W}.
\end{align}
It is known that $\Wc$ and $\Wm$ have a dipolar $r^{-3}$ decay at large
distance, see Section VI of \cite{buenzli-martin-ryser}.  One can therefore
view the system of loops as behaving like classical random charged wires
(interacting with $\Vel$) with additional electric and magnetic multipolar
interaction $\W$.

We deal with the screening effect generated by the classical Coulombic part
$\Vel$ by the standard Debye--H\"uckel method. This amounts to introduce
the effective screened potential $\Phi$ corresponding to the
chain-resummation of the linear part $-\beta e_{\gamma_i} e_{\gamma_j}
\Vel(\b i, \b j)$ of the bond $f(\b i, \b j)$: $\Phi$ satisfies the
integral equation
\begin{align}
    \Phi(\b i,\b j) = \Vel(\b i, \b j) - \int\!\!\!{\d \b 1}
    \frac{\kappa^2(1)}{4\pi} \Vel(\b i, \b 1) \Phi(\b 1, \b j),
    \label{Phiel-integral-def}
\end{align}
where
\begin{align}
    \kappa^{-1}(1) = [4\pi\beta e_{\gamma_1}^2 \rho(1)]^{-1/2}\label{kappa-loop}
\end{align}
defines a local screening length in the system of loops. This potential is
now short-range in the sense that it is integrable on the $\y$-direction
along the plates (see Appendix C), implying
\begin{align}
	\lim_{\k\to\0} |\Phi(1,2,\k)| < \infty. \label{phi-decay-prop}
\end{align}

The Mayer series is reorganised by the Abbe--Meeron resummation process
(\cite{brydges-martin}, Chap. V and references therein) into so-called
``prototype'' graphs $\Pi$ with integrable bonds $F(\b i, \b j)$ and
$\FR(\b i, \b j)$\footnote{At large distance, $\FR(\b i, \b j) \sim -\beta
e_{\gamma_i} e_{\gamma_j} \W(\b i, \b j) \sim r^{-3}$ is at the border of
integrability. Hence, some care has to be exercised as it is the case in
dipole gases.} given by
\begin{gather}
	F(\b i, \b j) = -\beta e_{\gamma_i} e_{\gamma_j} \Phi(\b i, \b j), \label{F}
	\\ \FR(\b i, \b j) = \e^{-\beta e_{\gamma_i}e_{\gamma_j} (\Phi+\W)(\b i,\b j)}-1+\beta e_{\gamma_i}e_{\gamma_j} \Phi(\b i, \b
        j). \label{FR}
\end{gather}
The resummed Mayer graph series of the Ursell function reads
\begin{align}
    h(\b 1, \b 2) = \sum_{\Pi} \frac{1}{S_\Pi} \int\!\!\!{\d\b 3}\,\rho(3) \cdots
    \!\!\!\int\!\!\!{\d\b m}\, \rho(m) \prod_{\{\b i, \b j\}\in\Pi} \F(\b i, \b j),
    \label{ursell-resummed-loop}
\end{align}
where $\F\in\{F, \FR\}$.  The diagrams $\Pi$ have two root points and $m-2$
internal circles (m=2,3,...), and a symmetry number $S_\Pi$. In
\eqref{ursell-resummed-loop}, the weights of the integrated points are the
density, so that the graphs contain no articulation points. Prototype
graphs are subject to an important rule: convolution chains of bonds $F$
are forbidden to avoid double counting of the original Mayer graphs.

\subsubsection*{Perfect screening sum rules}
On the microscopic level, the conducting behaviour of a system at
equilibrium is characterized by the fulfilment of the ``perfect screening
sum rule'' \cite{martin-sum-rules}: a fixed charge in the system is
neutralized by the mean charge density surrounding it. This property is
expressed by the following constraint on the two-particle Ursell function:
\begin{align}
	\sum_{\gamma_1}\int\!\! \d \r_1 \ e_{\gamma_1}\, \rho(\r_1,\gamma_1) h(\r_1,\gamma_1;
	\r_2,\gamma_2) = - e_{\gamma_2}. \label{perfect-screening}
\end{align}
It turns out that the same perfect screening sum rule holds in the
auxiliary system of loops
\begin{align}
	\int \!\! {\d \b 1}\ p_1\, e_{\gamma_1}\, \rho(\b 1) h(\b 1,\b 2) =
	-p_2\, e_{\gamma_2} \label{loop-perfect-screening}.
\end{align}
The interpretation is the same: the fixed loop $\b 2$ with charge $p_2
e_{\gamma_2}$ is surrounded by a screening cloud of loops with opposite
total charge.

The sum rule \eqref{loop-perfect-screening} holds in great generality for
infinitely extended conductors, in particular for slab geometries. A
justification of this sum rule is easily given when the loop Ursell
correlation $h$ is replaced by the single bond $F$. The equation
\eqref{Phiel-integral-def} written in the transverse Fourier space reads
\begin{align}
    \Phi(i,j, \k) = \Vel(i, j,\k) - \int\!\!\!{\d 1}
    \frac{\kappa^2(1)}{4\pi} \Vel(i, 1, \k) \Phi(1, j, \k),
    \label{Phiel-integral-k}
\end{align}
where from \eqref{Vel}
\begin{align}
    \Vel(i,j,\k) = \int_{0}^{p_i}\!\!\!\!\!{\d s_i} \!\!\int_{0}^{p_j}\!\!\!\!\!{\d
        s_j}\ \e^{i\k\cdot
        [\lambda_{\gamma_i}\Y_i(s_i)-\lambda_{\gamma_j}\Y_j(s_j)]}
    \,\tfrac{2\pi}{k}\e^{-k\left| x_i^{[s_i]}-x_j^{[s_j]}\right|} 
    \label{Vel-fourier}.
\end{align}
Here $X(s)$ and $\Y(s)$ are the components of $\X(s)$ along $x$ and in the
$\y$ plane and $x^{[s]}=x+\lambda_{\gamma}X(s)$ is the component along $x$
of $\r^{[s]}$ \eqref{loop-path}.  We divide both members of
\eqref{Phiel-integral-k} by $\Vel(i,j,\k)$ and let $\k\to 0$. In view of
the fact that $\Phi(i,j,\k)$ remains finite (see \eqref{phi-decay-prop}),
that \\$\lim_{\k\to\0} \Vel(i,1,\k)/\Vel(i,j,\k) = p_1/p_j$, and from the
definition \eqref{kappa-loop}, one obtains
\begin{align}
    \int\!\!{\d 1}\, p_1\, e_{\gamma_1}\, \rho(1)\, F(1,j,\k=\0) =
    -p_j e_{\gamma_j}\ ,
    \label{loop-DH-perfect-screening}
\end{align}
which is the same as \eqref{loop-perfect-screening} with $F$ replacing $h$.

In fact, it can be shown that the general case
\eqref{loop-perfect-screening} is a consequence of
\eqref{loop-DH-perfect-screening}, by using the same dressing argument as
that presented in (63)-(68) of \cite[Sec. 5]{buenzli-martin-classical}.

\section{Asymptotic Casimir force}
To analyse the asymptotic $d$-dependence of the force
\eqref{force-ursell-loop}, we need to extract that of the electrostatic
part $\p_{x_1} v_{AB}$ and of the magnetic part $\p_{x_1} \Wm_{AB}$
together with that of the Ursell correlation $h_{AB}$.  It is immediate
from \eqref{coulomb-force} that $\p_{x_1} v_{AB}(1,2,\tfrac{\q}{d})$ has
the limit
\begin{align}
	\p_{x_1} v_{AB}(1,2,\tfrac{\q}{d}) \to 2\pi \e^{-q}= \Order(1)
\end{align}
as $d\to\infty$. We will establish in Section
\ref{sec:asymptotic-correlations} the following facts:
\begin{align}
&\p_{x_1} \Wm_{AB}(1,2,\tfrac{\q}{d}) = \Order(d^{-2}),
\\&h_{AB}(1,2,\tfrac{\q}{d}) = \Order(d^{-1}).
\end{align}

As a consequence, the average of the magnetic part $\p_{x_1}
\Wm_{AB}(1,2,\tfrac{\q}{d})$ does not contribute to the Casimir force at
leading order since it is $\Order(d^{-5})$, whereas the electrostatic part
of the force is $\Order(d^{-3})$. To calculate the coefficient of this
$\propto d^{-3}$ dominant contribution, the exact structure of $h_{AB}$ at
$\Order(d^{-1})$ is needed. The latter is analysed in detail in Subsection
\ref{sec:hAB}. In short, both bonds $F_{AB}$ \eqref{F} and $\FR_{AB}$
\eqref{FR} are of order $d^{-1}$ and the diagrams contributing to
$h_{AB}(1,2,\tfrac{\q}{d})$ at this order comprise only one of these $AB$
links. Those having a single $F_{AB}$ bond sum up to the factorized
expression
\begin{align}
	    -\frac{1}{\beta d} \frac{q}{4\pi\sinh q}
        \frac{G_A^0(1,0,\0)}{e_{\alpha_0}}
        \frac{G_B^0(0,2,\0)}{e_{\beta_0}}
    \label{hAB-asymptotic-FAB},
\end{align}
where
\begin{align}
    G_A^0(1,0,\0) = &\ h_A^0(1,0,\0) \notag
    \\& - \int\!\!\!{\d i}\, \rho_A^0(i) \left[
        F_A^0(1,i,\0) + \tfrac{\delta(1,i)}{\rho_A^0(i)} \right]
    (h_A^0)^\text{nn}(i,0,\0) \label{GA0-hA0}
\end{align}
comprises internal correlations occurring in slab $A$. The superscript
``0'' qualifies statistical-mechanical quantities characterizing the system
governed by the same Hamiltonian \eqref{hamiltonian} but where
$V^\text{walls}$ confines particles in a single slab (either $A$ or
$B$). $h_A^0(1,0,\b 0) = h_A^0(1,0,\k=\0)$ is the Ursell correlation
between a loop ``$1$'' in slab $A$ and a classical charge $e_{\alpha_0}$
located at its right border, denoted by the loop argument
\begin{align}
    &0 \equiv \big(x\!=\!0,\,\alpha_0,\,p\!=\!1,\,\X(\cdot)\!\equiv\!
    \0\big). \label{loop-0}
\end{align}
The structure of $G_A^0$ is determined by the excluded convolution rule
applied to $F_{AB}$. The partial Ursell function $(h_A^0)^\text{nn}$
occurring in the right hand side of \eqref{GA0-hA0} is defined in
Subsection \ref{sec:hAB}, see \eqref{h-hnn}.  The same notations and
definitions apply to the plate $B$.

The diagrams having a single $\FR_{AB}$ bond sum up to the expression
\begin{align}
	&\int\!\!\!{\d i}\,\rho_A^0(i)
        \!\int\!\!\!{\d j}\,\rho_B^0(j) \left[h_A^0(1,i,\0) +
            \tfrac{\delta(1,i)}{\rho_A^0(i)} \right] \notag
    \\&\ \times(-\beta
    e_{\gamma_i} e_{\gamma_j}) \W_{AB}(i,j,\tfrac{\q}{d})
    \left[h_B^0(j,2,\0) + \tfrac{\delta(j,2)}{\rho_B^0(j)} \right]
    \label{hAB-asymptotic-WAB}
\end{align}
where again $h_A^0$ and $h_B^0$ are the Ursell functions of the single
plate systems $A$ and $B$.

The rest of the analysis relies on the application of the perfect sum rule
\eqref{loop-perfect-screening} for loops. Indeed, introducing the
contribution \eqref{hAB-asymptotic-WAB} into the force
\eqref{force-ursell-loop}, one builds the integral
\begin{gather}
	\int\!\! \d 1\ p_1 e_{\gamma_1} \rho_A^0(1) \left[h_A^0(1,i,\0) +
	\tfrac{\delta(1,i)}{\rho_A^0(i)} \right] = 0
\end{gather}
which vanishes by \eqref{loop-perfect-screening}.

Introducing now the contribution \eqref{hAB-asymptotic-FAB} into the force
\eqref{force-ursell-loop}, we see that the integrals on the two slabs
factorize as
\begin{align}
        f(d) &\stackrel{d\to\infty}{\sim}-\frac{1}{4\pi\beta d^3}
    \!\int_{0}^{\infty}\!\!\!\!{\d q}\ \frac{q^2 \e^{-q}}{\sinh q} 
    \notag
	\\&\phantom{\stackrel{d\to\infty}{\sim}}\times\!
    \left[\!\int\!\!{\d 1}\, p_1\, e_{\gamma_1}\,
        \rho_A^0(1)\tfrac{G_A^0(1,0,\0)}{e_{\alpha_0}}\right]
    \left[\!\int\!\!{\d 2}\, p_2\, e_{\gamma_2}\, \rho_B^0(2)
        \tfrac{G_B^0(0,2,\0)}{e_{\beta_0}}\right]. \label{force-FAB}
\end{align}

From \eqref{GA0-hA0}, we have
\begin{align}
\!\int\!\!{\d 1}\, p_1\, e_{\gamma_1}\,
        \rho_A^0(1)\tfrac{G_A^0(1,0,\0)}{e_{\alpha_0}} = \!\int\!\!{\d 1}\,
        p_1\, e_{\gamma_1}\, \rho_A^0(1)\tfrac{h_A^0(1,0,\0)}{e_{\alpha_0}}
        = -1.
\end{align}
The first equality follows from the sum rule
\eqref{loop-DH-perfect-screening} for the $F$ bond in the single plate
$A$. The second equality is again a consequence of the perfect screening
for loops \eqref{loop-perfect-screening}. Perfect screening in plate $B$
implies similar identities for the second bracket in \eqref{force-FAB}.

Noticing that the $q$-integral provides the constant $\zeta(3)/2$, the
Casimir force at large separation is
\begin{align}
        f(d) &\stackrel{d\to\infty}{\sim} - \frac{\zeta(3)}{8\pi\beta d^3}
        \label{force-result},
\end{align}
which is the main result of this paper.

\section{Asymptotic correlations between the two slabs}\label{sec:asymptotic-correlations}

To extract the asymptotic large-separation behaviour of the Ursell
correlation $h_{AB}(1,2,\tfrac{\q}{d})$, we select the class of prototype
graphs that give the dominant contribution by analysing them one by one, as
done in \cite[App. C]{buenzli-martin-classical}.

It is important to distinguish situations where arguments $\L_i, \L_j$ both
lie in the same plate or in the two different plates. As done before (see
\eqref{AB-notation}, \eqref{AB-notation2}) we index any quantity with
arguments $\L_i\in A=[0,a]\times\mathbb{R}^2$ and $\L_j\in
B=[0,b]\times\mathbb{R}^2$ with an index $AB$. We introduce a similar
notation for interactions and correlations internal to a given plate, using
the index $AA$, $BB$ when loops lie in the same slab, \eg,
\begin{align}
	&F_{AA} (\L_i,\L_j) = F(\L_i,\L_j), \quad F_{BB}(\L_i,\L_j) = F(\L_i+d, \L_j+d), \notag
	\\& h_{AA} (\L_i,\L_j) = h(\L_i,\L_j), \quad h_{BB}(\L_i,\L_j) = h(\L_i+d, \L_j+d).\label{AB-bonds}
\end{align}
In the limit $d\to\infty$, the plates will have no mutual interaction
anymore: $AB$-correlations are expected to vanish whereas $AA$ and $BB$
quantities will tend to those pertaining to the system constituted by a
single plate. Using the superscript ``$0$" to qualify the statistical
mechanical description of the single plates, one will have in particular
\begin{gather}
    \rho_A(1) \stackrel{d\to\infty}{\longrightarrow} \rho_A^0(1),
    \qquad \rho_B(2) \stackrel{d\to\infty}{\longrightarrow} \rho_B^0(2),
   \notag
    \\ F_{AA}(\b i, \b j) \to F_A^0(\b i, \b j), \qquad F_{BB}(\b i, \b j) \to F_B^0(\b i, \b j), 
     \label{quantities-to-isolated}
    \\h_{AA}(\b i, \b j) \to h_A^0(\b i, \b j), \qquad h_{BB}(\b i, \b j) \to h_B^0(\b i, \b j).
    \notag
\end{gather}
In the next subsections, we analyse in more detail the behaviour of
$h_{AB}(1,2,\tfrac{\q}{d})$ occurring in the force formula
\eqref{force-ursell-loop} at large $d$.

\subsection{Large-distance behaviour of the screened potential $\Phi_{AB}$}\label{sec:phiAB}
The main fact to be established in this subsection is the factorization of
the screened potential $\Phi_{AB}(i,j,\tfrac{\q}{d})$, at order $d^{-1}$, into two
independent parts pertaining to the individual slabs $A$ and $B$. 
We extend the arguments developed for a system of classical charges presented in Section 3.2.3 of Ref. \cite{martin-buenzli-kazimierz}. One observes first that this factorization is already present in the bare Coulomb
potential $ \Vel(i, j, \k)$ \eqref{Vel-fourier}. Indeed taking into
account the shift \eqref{shift} of the loops' positions as well as $
x_i^{[s_i]} < 0 < x_j^{[s_j]}$ for all $s_i,s_j$, one can write
\begin{align}
    \Vel_{AB}(i,j,\k) &= \frac{k \e^{-kd}}{2\pi} \Bigg[\int_{0}^{p_i}\!\!\!{\d
            s_i}\ \e^{i\k\cdot\lambda_{\gamma_i} \Y_i(s_i)} \tfrac{2\pi}{k}
        \e^{-k \big|x_i^{[s_i]}\big|} \Bigg]\notag
    \\&\phantom{=}\times
    \Bigg[\int_{0}^{p_j}\!\!\!{\d
            s_j}\ \e^{-i \k \cdot \lambda_{\gamma_j}\Y_j(s_j)}
        \tfrac{2\pi}{k} \e^{-k \big|x_j^{[s_j]}\big|} \Bigg] \notag
    \\&\equiv \frac{k \e^{-kd}}{2\pi}\ \Vel_{AA}(i, 0, \k) \ \Vel_{BB} (0,j,\k).
    \label{VelAB-factorization}
\end{align}
We have identified the first bracket to the Coulomb potential inside $A$
between a loop $i$ and a loop variable denoted $0$, corresponding to an
(arbitrary) classical charge situated on the inner side of the slab (see
\eqref{loop-0}), and likewise for the second bracket. We show below that
the factorization extends to the screened potential as $d\to\infty$ in the
form
\begin{align}
    \Phi_{AB}(i,j,\tfrac{\q}{d}) \stackrel{d\to\infty}{\sim} &\frac{1}{d}
    \frac{q}{4\pi \sinh q}\ \Phi_A^0(i,0,\0)\,
    \Phi_B^0(0,j,\0), \label{PhiAB-asymptotic}
\end{align}
where $ \Phi_A^0(i,0,\0)$ is the screened potential at $\k=\0$ inside the
single plate $A$ between a loop $i$ and a classical charge $0$ at its right
boundary, and likewise for $\Phi_B^0(0,j,\0)$.

In the chain summation of $\Vel$ bonds that constitutes $\Phi_{AB}$,
obtained by iterating \eqref{Phiel-integral-k}, we keep only the dominant
chains, which turn out to be of order $d^{-1}$. We follow the steps
performed in the classical situation in \cite{martin-buenzli-kazimierz}. We
split every integral on internal convolution points into an $A$ and a $B$
contribution. Using again the notation \eqref{AB-bonds}, i.e. specifying
$\Vel$ as $\Vel_{AA}(i,j,\k), \Vel_{BB}(i,j,\k), \Vel_{AB}(i,j,\k) $
according to the location of its argument in $A$ or $B$, $\Vel$ chains are
expanded into chains made of $\Vel_{AA},\Vel_{BB}$ and $\Vel_{AB}$
bonds. One notes that $\Vel_{AA}(i,j,\k)=\Vel_{BB}(i,j,\k)$, and that
$\Vel_{BA}(i,j,\k)=\Vel_{AB}(j,i,-\k)$ by space inversion in the $\y$
plane. We call $\Vel_{AB}(i,j,\k)$ a traversing bond, and chains that link
$A$ with $B$, traversing chains. Clearly, traversing chains that contribute
to $\Phi_{AB}(i,j,\tfrac{\q}{d})$ have necessarily an odd number of
traversing bonds $\Vel_{AB}$. Let $\Phi_{AB}^{(2n+1)}$ be the sum of chains
containing exactly $2n+1$ traversing bonds. The contribution
$\Phi_{AB}^{(1)}$ is a sum of convolution chains of the type $\cdots
\Vel_{AA}\ast \Vel_{AA} \ast\Vel_{AB}\ast \Vel_{BB} \ast\Vel_{BB}
\cdots$. Using the factorization \eqref{VelAB-factorization} of
$\Vel_{AB}$, one can resum on eiter side of $\Vel_{AB}$ convolution chains
of $\Vel_{AA}$ and $\Vel_{BB}$ into quantities
$\widetilde\Phi_{AA}(i,0,\tfrac{\q}{d})$ and
$\widetilde\Phi_{BB}(0,j,\tfrac{\q}{d})$; $\widetilde\Phi_{AA}$ differs
from the screened potential $\Phi_{AA}$ by the omission of traversing
chains starting in plate $A$ and returning to it, which describe part of
the electrical influence of $B$ on $A$ (likewise for
$\widetilde\Phi_{BB}$). But for large $d$, these traversing chains do not
contribute anymore and by \eqref{quantities-to-isolated},
$\widetilde\Phi_{AA}(i,0,\tfrac{\q}{d})$ and
$\widetilde\Phi_{BB}(0,j,\tfrac{\q}{d})$ also tend as $d\to\infty$ to the
screened loop potentials $ \Phi_A^0(i,0,\0)$ and $ \Phi_B^0(0,j,\0)$ of the
single plates systems.  Hence,
\begin{gather}
    \Phi^{(1)}_{AB}(i,j,\tfrac{\q}{d}) \stackrel{d\to\infty}{\sim} \frac{q
        \e^{-q}}{2\pi d}\ {\Phi}_{A}^0(i, 0,
    \0)\ {\Phi}_{B}^0(0, j,
    \0) \label{PhiAB1-factorization}
\end{gather}
is of order $d^{-1}$ with a factorized coefficient.

The contribution $\Phi_{AB}^{(3)}$ is then formed by convolution chains of
the type $\cdots \Vel_{AA}\ast \Vel_{AB}\ast \Phi_{BA}^\text{(1)}\ast \Vel_{AB}\ast
\Vel_{BB}\cdots$. Using \eqref{VelAB-factorization},
\eqref{PhiAB1-factorization} and resumming again the convolution chains of
$\Vel_{AA}$ and $\Vel_{BB}$ on either extremity into $\widetilde\Phi_{AA}$
and $\widetilde\Phi_{BB}$, one obtains
\begin{align}
    \Phi^{(3)}_{AB}(i,j,\tfrac{\q}{d}) = &\left(\frac{q \e^{-q}}{2\pi
        d}\right)^3 \widetilde{\Phi}_{AA}(i,0,\tfrac{\q}{d}) \left[ -\!\!\int\!\!{\d
            1} \frac{\kappa_B^2(1)}{4\pi} \Vel_{BB}(0,1,\tfrac{\q}{d})
        \widetilde{\Phi}_{BB}(1,0,\tfrac{\q}{d}) \right] \notag
    \\&\times \left[ -\!\!\int\!\!{\d 2} \frac{\kappa_A^2(2)}{4\pi}
        \widetilde{\Phi}_{AA}(0,2,\tfrac{\q}{d}) \Vel_{AA}(2,0,\tfrac{\q}{d})
        \right] \widetilde{\Phi}_{BB}(0,j,\tfrac{\q}{d}).
    \label{Phi3}
\end{align}
By definition of $\widetilde\Phi_{AA}$ and $\widetilde\Phi_{BB}$, these
quantities satisfy the integral relation \eqref{Phiel-integral-k} relative
to $A$ and $B$ with inverse screening lengths $\kappa_A^2(1)$ and
$\kappa_B^2(1)$ in place of $\kappa^2(1)$. The brackets in \eqref{Phi3}
thus reduce to
\begin{align}
    &\widetilde\Phi_{BB}(0,0,\tfrac{\q}{d}) - \Vel_{BB}(0,0,\tfrac{\q}{d}) =
    -\frac{2\pi d}{q} + \Order(1), \label{vBB-PhiBB}
    \\&\widetilde\Phi_{AA}(0,0,\tfrac{\q}{d}) - \Vel_{AA}(0,0,\tfrac{\q}{d}) =
    -\frac{2\pi d}{q} + \Order(1). \label{vAA-PhiAA}
\end{align}
On the right hand side of \eqref{vBB-PhiBB} and \eqref{vAA-PhiAA}, the
dominant terms come from the Coulomb potentials (see \eqref{Vel-fourier}),
while the estimates $\Order(1)$ reflect the fact that
$\widetilde\Phi_{AA}(i,j,\k)$ and $\widetilde\Phi_{BB}(i,j,\k)$ are bounded
in $\k$ (see \eqref{phi-decay-prop} and Appendix C) . This yields
\begin{align}
    \Phi^{(3)}_{AB}(i,j,\tfrac{\q}{d})\ \stackrel{d\to\infty}{\sim}\ 
    &\frac{q\e^{-q}}{2\pi d}\, \e^{-2q} {\Phi}_{A}^0(i,0,\0)
    {\Phi}_{B}^0(0,j,\0). \label{PhiAB3-asymptotic}
\end{align}
By induction on $n$, one easily sees that
$\Phi^{(2n+1)}_{AB}(i,j,\tfrac{\q}{d})$ receives instead a prefactor
$\frac{q\e^{-q}}{2\pi d} \e^{-2nq}$. Summing over $n=1,2,3,...$ gives the
result \eqref{PhiAB-asymptotic}.  It is interesting to see that the
screened electrostatic interaction between the plates, at order $d^{-1}$,
involves only particles located close to the inner faces of the slabs.

For later convenience we write the equivalent factorized form of the bond
$F$ \eqref{F},
\begin{align}
    F_{AB}(i,j,\tfrac{\q}{d}) \stackrel{d\to\infty}{\sim} &- \frac{1}{\beta d}
    \frac{q}{4\pi \sinh q}\ \frac{F_A^0(i,0,\0)}{e_{\alpha_0}}\,
    \frac{F_B^0(0,j,\0)}{e_{\beta_0}}, \label{FAB-asymptotic}
\end{align}
where $e_{\alpha_0}$ and $e_{\beta_0}$ are two charges located at the inner
boundary of the slabs.

\subsection{Large-distance behaviour of the dipolar potential $\W_{AB}$}\label{sec:WAB}
The partial Fourier transform $\W_{AB}(1,2,\k)$ is related to the
three-dimensional Fourier transform $\W(\chi_1,\chi_2, \K)$ ($\K=(k_1,\k)$)
of \eqref{W} by
\begin{align}
	\W_{AB}(1,2,\k) = \int\!\! \frac{\d k_1}{2\pi} \e^{i k_1(x_1-x_2-d)}
	\W(\chi_1,\chi_2,\K), \label{WAB-fourier1D}
\end{align}
remembering that $\W_{AB}(\L_1,\L_2) \equiv \W(\L_1,\L_2+d)$. Changing
$k_1\mapsto q_1/d$ and setting $\k=\q/d$,
\begin{align}
	\W_{AB}(1,2,\tfrac{\q}{d}) = \frac{1}{d} \int\!\! \frac{\d q_1}{2\pi}
	\e^{i q_1\frac{x_1-x_2}{d}} \e^{-i q_1}
	\W(\chi_1,\chi_2,\tfrac{q_1}{d}, \tfrac{\q}{d}), \label{WAB-Fourier3D}
\end{align}
which shows that $\W_{AB}(1,2,\tfrac{\q}{d})=\Order(d^{-1})$ provided the
integral has a limit as $d\to\infty$. The analysis of
$\W(\chi_1,\chi_2,\K)$ at small $\K$ has been carried out in
\cite{buenzli-martin-ryser}. It was observed that the dipolar electric part
$\Wc$ was screened by thermalized photons at large distance. As a
consequence $\W(\chi_1,\chi_2,\K)$ behaves as
$W^\text{m}(\chi_1,\chi_2,\K)$ when $\K\to 0$, where $W^\text{m}$ is the
magnetic potential corresponding to a classical electromagnetic field (\ie,
setting $\lambda_\text{ph}\equiv 0$ in $\Wm$). The behaviour of
$W^\text{m}(\chi_1,\chi_2,\K)$ itself was worked out in
\cite[Form. (25)]{boustani-buenzli-martin}:
\begin{align}
	W^\text{m} (\chi_1,\chi_2,\K) &\sim
	\frac{\lambda_{\gamma_1}\lambda_{\gamma_2}}{\beta
	\sqrt{m_{\gamma_1}m_{\gamma_2}} c^2} \int_0^{p_1}\!\!\!\d X_1^\mu(s_1)
	\!\!  \int_0^{p_2}\!\!\!\d X_2^\mu(s_2) \notag
	\\&
	\times [\K \cdot \X_1(s_1)] [\K \cdot \X_2(s_2)] \frac{4\pi}{\nK^2}
	\delta_{\mu\nu}^\text{tr}(\K), \quad \K\to\0. \label{Wm-class-asymptotic}
\end{align}
It is analogous to dipolar magnetic interaction between two classical
current loops of shape $\X_1(\cdot)$ and $\X_2(\cdot)$. Since this
interaction only depends on the unit vector $\hat\K=\K/\nK$,
$\W(\chi_1,\chi_2, \tfrac{q_1}{d}, \tfrac{\q}{d})$ is asymptotically
independent of $d$, implying
\begin{align}
	\W_{AB}(1,2,\tfrac{\q}{d})=\Order(d^{-1}) \label{WAB-asymptotic}.
\end{align}
An explicit expression for the asymptotic form of \eqref{WAB-Fourier3D} can
be found in Appendix \ref{appx:Wm-asymptotic}.

The potentials $\Wc_{AB}$ and $\Wm_{AB}$ could as well be separately
analysed in the same way. One sees on \eqref{WAB-Fourier3D} that $\p_{x_1}
\Wm_{AB}(1,2,\tfrac{\q}{d})$ has an additional $d^{-1}$ factor, so that
\begin{align}
	\Wm_{AB}(1,2,\tfrac{\q}{d}) = \Order(d^{-1}), \quad \p_{x_1} \Wm_{AB}
	(1,2,\tfrac{\q}{d}) = \Order(d^{-2}). \label{magnetic-force}
\end{align}

\subsection{The Ursell function at order $\Order(d^{-1})$}\label{sec:hAB}
From \eqref{FAB-asymptotic}, \eqref{WAB-asymptotic} and the definition of
$\FR_{AB}$ \eqref{FR}, one sees that $F_{AB}(i,j,\tfrac{\q}{d}) =
\Order(d^{-1})$ and $\FR_{AB}(i,j,\tfrac{\q}{d}) = \Order(d^{-1})$.  It is
clear that the decay rate of a prototype graph will depend on the number of
its traversing bonds. A rough counting gives $d^{-n_{AB}}
d^{-n^\text{R}_{AB}}$ where $n_{AB}$ is the number of $F_{AB}$ bonds and
$n^\text{R}_{AB}$ the number of $\FR_{AB}$ bonds. A closer inspection shows
that this decay can be even faster, at least as
\begin{align}
	d^{-2I} d^{-n_{AB}} d^{-n^\text{R}_{AB}}, \qquad d\to\infty,
	\label{graph-decay}
\end{align}
where $I$, $0\leq I \leq n_{AB}+n^\text{R}_{AB}-1$, depends on the topology
of the specific diagram and $I=0$ if there is a single traversing
bond. Formula \eqref{graph-decay} can be established repeating the same
steps as Appendix C of \cite{buenzli-martin-classical} (with integrals over
loop degrees of freedom).

From \eqref{graph-decay}, the slowest decaying Mayer diagrams have either
$n_{AB}=1, n^\text{R}_{AB}=0$ or $n_{AB}=0, n^\text{R}_{AB}=1$, namely only
one $F_{AB}$ or one $\FR_{AB}$ bond. In forming the complete correlation
function of the two-slab system, these bonds have to be dressed at their
extremities by appropriate internal correlations of the individual slabs in
conformity with the diagrammatic rules. Thus, the complete expression of
the Ursell function $h_{AB}(1,2,\tfrac{\q}{d})$ at order $\Order(d^{-1})$
is
\begin{align}
	h_{AB} \stackrel{d\to\infty}{\sim} D_{AA}\ast F_{AB} \ast D_{BB} +
	D_{AA}^R \ast \FR_{AB}\ast D_{BB}^R.
\end{align}
The formation of the dressing function $D$ differs from that of $D^R$
because of the excluded convolution rule in prototype graphs: no $F_{AA}$
or $F_{BB}$ bond can be attached alone to the extremities of $F_{AB}$
whereas there is no such restriction for $\FR_{AB}$.

The dressing function $D_{AA}^R$ ($D_{BB}^R$) of $\FR_{AB}$ consists of all
possible $AA$ ($BB$) internal graphs. According to the discussion at the
beginning of the section (see \eqref{quantities-to-isolated}), it tends to
the Ursell function $h_A^0$ ($h_B^0$) of the individual plate.  The
corresponding contribution to $h_{AB}(1,2,\tfrac{\q}{d})$ is thus given at
$\Order(d^{-1})$ by \eqref{hAB-asymptotic-WAB}. The delta terms in
\eqref{hAB-asymptotic-WAB} account for the situation where no bonds are
attached to the extremities of $\FR_{AB}$.

To deal with the excluded convolution rule when forming the dressing
function $D_{AA}$, we introduce the function $h_{AA}^{nn}(\b i, \b j)$
defined by the sum of all prototype graphs that do not begin nor end with a
$F$ link alone. Its relation to the Ursell function is
\begin{align}
h = F + h^{nn}\ast F + F \ast h^{nn}\ast F + F \ast h^{nn} + h^{nn} \label{h-hnn}.
\end{align}
Then $D_{AA} = h_{AA}^{nn} + F_{AA}\ast h_{AA}^{nn} + \delta/\rho_A$
(likewise for $D_{BB}$). With the factorization \eqref{FAB-asymptotic} of
the link $F_{AB}$, one sees that $(D_{AA}\ast F_{AB} \ast
D_{BB})(1,2,\tfrac{\q}{d})$ has the factorized form
\eqref{hAB-asymptotic-FAB} with
\begin{align}
	G_A^0 = \left( \frac{\delta}{\rho_A^0} + (h_A^0)^{nn} +F_A^0\ast
	(h_A^0)^{nn} \right) \ast F_A^0. \label{GA0}
\end{align}
Since the bond $F_{AB}$ is already $\Order(d^{-1})$, all other quantities
have been evaluated for single plate systems according to
\eqref{quantities-to-isolated}. The final form \eqref{GA0-hA0} of $G_A^0$
follows by noticing that the first three terms of \eqref{h-hnn} build
$G_A^0$ as given in \eqref{GA0}.

\section{Concluding remarks}

In this paper the large-separation asymptotics \eqref{force8pibeta} of the
Casimir force between two conducting plates has been derived exactly from
the principles of quantum electrodynamics and statistical mechanics for any
fixed nonzero temperature, taking all microscopic degrees of freedom of
matter and field into account. This does not give a direct proof that the
TE mode reflexion coefficient does not contribute at zero frequency, but a
strong evidence for it. The derivation applies to any model of conductor
consisting of mobile quantum charges. The latter can be negative and
positive charge carriers (like ions and anions in electrolytes), or, \eg,
form the one-component electron gas in the jellium model of a metal,
the central common point to all these systems being the screening
mechanisms and the perfect screening sum rules.

Let us note that there can be no contradiction between the behaviour
\eqref{force8pibeta} and the Nernst heat theorem. As mentioned in the
introduction, it has been argued, and controversially debated, that the use
of the Drude expression of the dielectric function (yielding
\eqref{force8pibeta}) was not consistent with the Nernst postulate which
requires that the entropy of the total system vanishes at zero
temperature. In our setting the question arises in different terms. The
asymptotic formula \eqref{force8pibeta} is definitely true, whereas the
Nernst theorem is the separate affirmation that the QED Hamiltonian
\eqref{hamiltonian} has a unique (or not extensively degenerate) ground
state, a nontrivial and uncorrelated mathematical problem.

A number of questions deserve further studies. We have disregarded
paramagnetic forces due to the Pauli coupling $-\mu \b\sigma\! \cdot\! {\b
B }$ of electronic and nuclear magnetic moments $\mu \b \sigma$ to the
magnetic field ${\b B }$. Preliminary investigations using spin coherent
states functional integrals indicate that such interactions result in an
additional effective dipolar potential which, as the orbital diamagnetic
part $\Wm$, will not contribute to the asymptotic force.

We have kept the thickness $a$ and $b$ of the plates finite while the
separation $d\to\infty$. Then, because of perfect screening, the asymptotic
force is universal as well as independent of $a$ and $b$. This corresponds
to the present experimental situation where only thin coats of metal of
order of $50$ nm are deposited on a substrate \cite{bressi-etal}. Compared
to separations ranging from $0.5$ to $3$ $\mu$m, the regime is clearly
$a,b\ll d$. The opposite situation of thick plates $a,b\gg d$, namely
taking here $a=b=\infty$ at the very beginning, requires a careful analysis
since the magnetic potential $\Wm(1,2,\tfrac{\q}{d})$ looses integrability
as $d\to\infty$ (see the factor $\exp(i\, q_1\, [x_1\!-\!x_2]/d)$ in
\eqref{WAB-Fourier3D}). Then $x$-integrals have to be performed before
taking the limit $d\to\infty$ which appears to lead to a modification of
the $d^{-3}$ coefficient with small (nonuniversal) terms of order
$\Order\big((\beta m c^2)^{-1})\big)$.

Expression \eqref{force-result} is the first term of an expansion in
inverse powers of $d$ whose terms will be of the form $A_n/d^n,\;n\geq
4$. The amplitudes $A_n(\rho,T)$ are no more universal. They will depend on
the thermodynamic and geometric parameters of the plates (temperature,
densities $\rho$, thickness $a$, $b$) as well as their microscopic
characteristics (particle masses and charges). Looking at the form of the
electrostatic and magnetic dipole potentials $\Wm$ and $\Wc$, the expansion
can be cast in terms of dimensionless quantities including
$(\lambda_\text{mat}/d)^{n},\,(\lambda_\text{ph}/d)^{n}$, where
$\lambda_\text{mat}$ and $\lambda_\text{ph}$ are the matter and photon
thermal lengths. This expansion is therefore only meaningful when the
condition \eqref{ineq1} is met. Of great interest would be the calculation
of the first subdominant amplitude $A_4$ that includes corrections from an
imperfectly-conducting metal and to compare it with the predictions of the
Lifshitz theory \cite{milton-controversies-progress}. Also the effect of
the capacitor force, which cannot be completely turned off in experiment,
can be estimated by analysing the term \eqref{capacitor-force} at large
separation.
 
Finally, an open question is the understanding of the crossover to the
zero-temperature Casimir force $f\sim -\pi^2\hbar c/240 d^4$ due to pure
quantum fluctuations.
In the framework of Lifshitz theory, it was shown that plasmon modes at the
surfaces of the plates combine with photonic modes to build the above usual
zero-temperature Casimir force calculated as if the plates were inert
\cite{intravaia-lambrecht}. Notice that we have not added to the
Hamiltonian \eqref{hamiltonian} the vacuum energy
$\tfrac{1}{2}\sum_{K,\lambda}\hbar\omega_K$. In fact this (infinite)
constant plays no role since it will anyway not appear in the force formula
\eqref{casimir-force-def} (it is independent of $d$ in our setting). To
study the zero-temperature case one cannot rely on the above-mentioned
expansion since $\lambda_\text{mat}, \lambda_\text{ph}\to \infty$ as $T\to
0$ and condition \eqref{ineq1} does not hold anymore.  One has to
reconsider the whole analysis by first evaluating the force
\eqref{force-ursell-loop} in the zero-temperature limit at fixed $d$ and
then let $d\to\infty$. In other words, the limits $T\to 0$ and $d\to\infty$
are not permutable, and the issue is about obtaining a simultaneous control
of the force jointly for $T$ near zero and $d$ large. This will be the
subject of forthcoming work.

\appendix

\section{Capacitor force}\label{appx:capacitor-force}

Given that in the electrostatic part of the total force \eqref{force-loops-y}, only
the monopolar part $p_1 p_2 \p_{x_1} v_{AB}$ of the loop Coulomb force
$\p_{x_1}\Vc_{AB}$ contributes (see Appendix \ref{appx:electro-force}), the
electrostatic capacitor force $\fc_\text{cap}(d)$ is
\begin{align}
	\fc_\text{cap}(d) = \int_A\!\!\!\d 1 \!\!\int_B\!\!\!\d 2 \!\!
	\int\!\!\!\d\y\,
	e_{\gamma_1}e_{\gamma_2}\, p_1 p_2\,\p_{x_1}v_{AB}(1,2,\y) \rho_A(1)\rho_B(2).
\end{align}
The loop densities $\rho_A(1)$ and $\rho_B(2)$ are independent of $\y$ by
space homogeneity in the plates' directions and $\int\!\!\d \y\
\p_{x_1}v_{AB}(1,2,\y)=2\pi$ (set $\q=\0$ in \eqref{coulomb-force}). The remaining
integrals factorize, and yield the particle densities
$\rho_A(x_1,\gamma_1)$ and $\rho_B(x_2,\gamma_2)$ in plate $A$ and $B$ by
means of the identity
\begin{align}
	\rho(\r,\gamma) = \sum_{p=1}^\infty p \int\!\!\D(\X)\ \rho(\L)
\end{align}
(see \cite[Appendix D]{ballenegger-etal}). Introducing the charge density
$c(x) = \sum_{\gamma} e_{\gamma} \rho(x,\gamma)$ in plate $A$ and $B$,
$\fc_\text{cap}(d)$ is thus given by Formula \eqref{capacitor-force}. Note that the
charge densities $c_A(x)$ and $c_B(x)$ are still subject to the mutual
interaction between the slabs, thus depend on the separation $d$.

The magnetic part of the capacitor force is
\begin{align}
	\fm_\text{cap}(d) = \int_A\!\!\!\d 1 \!\!\int_B\!\!\!\d 2 \!\!
	\int\!\!\!\d\y\,
	e_{\gamma_1}e_{\gamma_2}
	\p_{x_1}\Wm_{AB}(1,2,\y) \rho_A(1)\rho_B(2). \label{magnetic-capacitor-force}
\end{align}
We show hereafter that
\begin{align}
	\int\!\!\d\y\, \p_{x_1}\Wm_{AB}(1,2,\y) = \int\!\!\frac{\d k_1}{2\pi}\, \e^{i
	k_1(x_1-x_2-d)} \, i k_1 \, \Wm(\chi_1,\chi_2,k_1,\k=\0) \label{magnetic-force-zero-k}
\end{align}
decays faster than any inverse power of
$(x_1-x_2-d)$ as $d\to\infty$. This ensures that the decay of
$\fm_\text{cap}(d)$ with the plates' separation has no power-law tail in
view of \eqref{magnetic-capacitor-force}.

The dipolar decay of $\Wm(\b 1, \b 2)$ \eqref{Wm} at large distance, responsible for
the power-law estimates \eqref{magnetic-force}, is generated by the
nonanalyticity $k_\mu k_\nu/\nK^2$ due to $\delta_{\mu\nu}^\text{tr}(\K)$
in the transverse Coulomb potential $4\pi
\delta_{\mu\nu}^\text{tr}(\K)/\nK^2$ ($\K=(k_1,\k)$). However, setting
$\k=\0$ eliminates the $k_1$ dependency in the transverse Kronecker symbol:
\begin{align}
	\delta_{\mu\nu}^\text{tr}(k_1,\k\!=\!\0) = \begin{cases}
		\delta_{\mu\nu} \quad &\text{if}\ \mu,\nu\neq 1,
		\\0 \quad &\text{if}\ \mu=\nu=1.
		\end{cases}
\end{align}
Any nonanalyticity is thus removed in $\Wm(\chi_1,\chi_2,k_1,\k\!=\!\0)$
around $k_1\!=\!0$, ensuring the fast decay of
\eqref{magnetic-force-zero-k}. Indeed, in
$\Wm(\chi_1,\chi_2,k_1,\k\!=\!\0)$, one is left with the stochastic
integrals
\begin{align}
	\int_0^{p_1}\!\!\!\!\!\d\Y_1(s_1) \cdot
	\!\!\int_0^{p_2}\!\!\!\!\!\d\Y_2(s_2)\notag \frac{4\pi
	g^2(k_1)}{k_1^2}\mathcal{Q}(k_1,\widetilde{s_1}\!-\!\widetilde{s_2})
	\e^{i k_1 \lambda_{\gamma_1} X_1(s_1)} \e^{-i k_1 \lambda_{\gamma_2}
	X_2(s_2)}.
\end{align}
Both $g^2(k_1)$ and $\mathcal{Q}(k_1,\widetilde{s_1}\!-\!\widetilde{s_2})$
are analytic functions of $k_1$ expandable as $1+\Order(k_1^2)$. One sees
by expanding the integrant around $k_1=0$ that the only singular terms are
functions of only $s_1$ or $s_2$. Their stochastic integration identically
vanishes by It\^o's lemma, stating that $\int_0^p \d\X(s) \equiv 0$.

\section{Electrostatic force}\label{appx:electro-force}

In Formula \eqref{force-average-loops}, the Casimir force has an
electrostatic part $\fc(d)$ due to $\p_{x}\Vc_{AB}$ and a magnetic part
$\fm(d)$ resulting from differentiating the magnetic potential. One could
write the average electrostatic force between the two slabs directly by
summing the Coulomb forces between the point charges:
\begin{align}
	\fc(d) &= \lim_{L\to\infty} \frac{1}{L^2} \avg{\sum_a\sum_b e_{\gamma_a}
	e_{\gamma_b} \p_x v_{AB}(\r_a, \r_b)} \notag
	\\&= \lim_{L\to\infty} \frac{1}{L^2}
	\!\!\int_{\! A}\!\!\!\d\r_1\!\!\int_{\! B}\!\!\!\d\r_2
	\sum_{\gamma_1}\sum_{\gamma_2} e_{\gamma_1} e_{\gamma_2} \p_x
	v_{AB}(\r_1,\r_2) \notag
	\\ &\phantom{=} \times \rho^{(2)}_{AB,
	L}(\r_1, \gamma_1;\r_2, \gamma_2), \label{point-electro-force}
\end{align}
where $v_{AB}(\r_a,\r_b)$ is the Coulomb potential \eqref{v},
$\rho^{(2)}_L$ is the particle density correlation function, and the same
notation \eqref{AB-notation}, \eqref{AB-notation2} translating positions
from slab $B_d$ to slab $B$ is used. Going to the phase space of loops by
means of the identity
\begin{align}
	\rho^{(2)}_{AB,L}(\r_1,\gamma_1;\r_2,\gamma_2) = \sum_{p_1=1}^\infty
	\sum_{p_2=1}^\infty \! p_1 p_2 \!\int\!\!\!\D(\X_1) \!\! \int
	\!\!\!\D(\X_2) \rho^{(2)}_{AB,L}(\b 1,\b 2)
\end{align}
(see \cite[Appendix D]{ballenegger-etal}) yields
\begin{align}
	\fc(d) =  \lim_{L\to\infty} \frac{1}{L^2} \int_{\! A}\!\!\d\b 1 \!\!
	\int_{\! B}\!\!\d\b 2\ 
	e_{\gamma_1}e_{\gamma_2} \big(p_1\, p_2\, \p_{x_1}v_{AB}\big)(\b 1,\b 2)
	\rho^{(2)}_{AB,L}(\b 1, \b 2).\label{electro-force-simplified}
\end{align}
In Formula \eqref{electro-force-simplified}, multipolar contributions
of the Coulomb force are not present, in contrast to the electrostatic part
of \eqref{loop-micro-force-c-m}. The strict equivalence of these formulae
relies on an invariance property regarding the choice of a reference point
for a loop's position. Clearly, the loop
\begin{align}
	\L^{[u]} \equiv \big(\r^{[u]}, \gamma, p, \X^{[u]}(\cdot)\big) \quad
	\text{with}\quad  \X^{[u]}(s) \equiv \X(s+u)-\X(u)
\end{align}
describes the same path as the loop $\L=\big(\r,\gamma,p,\X(\cdot)\big)$
but has its origin $\r^{[u]}=\r+\lambda_\gamma \X(u)$ shifted by the vector
$\lambda_\gamma \X(u)$ (the time parameter is shifted by $u$). Such a shift
does not affect the loop density:
\begin{align}
	\rho(\L^{[u]}) = \rho(\L) \quad \forall u, \label{density-invariance}
\end{align}
whereas the two-loop correlation function satisfies
\begin{align}
	\rho^{(2)}_{L}(\L_1^{[u_1]},\L_2^{[u_2]}) =
	\rho^{(2)}_{L}(\L_1,\L_2) \quad \text{if}\quad u_1-u_2 \in \mathbb{Z}
	\label{correlation-invariance}
\end{align}
(see below). In the electrostatic part of \eqref{loop-micro-force-c-m},
\begin{align}
	(\p_{x_1} \Vc_{AB})(\b 1,\b 2) = \equaltime{p_1}{s_1}{p_2}{s_2} \p_{x_1}
	v_{AB}(\r_1^{[s_1]},\r_2^{[s_2]}) \label{loop-coulomb-force}
\end{align}
(see \eqref{Vc}) and one can replace $\rho^{(2)}_{AB,L}(\L_1,\L_2)$ by
$\rho^{(2)}_{AB,L}(\L_1^{[s_1]},\L_2^{[s_2]})$ (at fixed $p_1, p_2, s_1,
s_2$) because of the equal-time constraint in \eqref{loop-coulomb-force}
that forces $s_1-s_2$ to be integer. Performing first the changes of
variable $\r_1\mapsto \r_1^{[s_1]}$, $\r_2\mapsto\r_2^{[s_2]}$ and then
\begin{align}
	\X_1(\cdot) \mapsto \X_1^{[s_1]}(\cdot), \quad \X_2(\cdot) \mapsto
	\X_2^{[s_2]}(\cdot),
	\label{X-change}
\end{align}
one obtains the electrostatic force
\begin{align}
	\lim_{L\to\infty}\!\frac{1}{L^2}\!\!\int_{\! A}\!\!\! \d \b 1
	\!\!\int_{\! B} \!\!\! \d \b 2
	\, e_{\gamma_1}e_{\gamma_2} \!\!\equaltime{p_1}{s_1}{p_2}{s_2}\!
	\p_{x_1}v_{AB}(\r_1,\r_2) \rho^{(2)}_{AB,L}(\b 1, \b 2)
	\label{electro-force-almost-simplified}.
\end{align}
Indeed, the Jacobian of the transformations \eqref{X-change} is equal to
$1$: the random process $\X^{[u]}(\cdot)$ is still Gaussian with unit
normalization, zero mean, and same covariance \eqref{X-covariance} as
$\X(\cdot)$, so that the Gaussian measure is unchanged:
$\D(\X^{[u]})=\D(\X)$ (see \cite[Lemma 1]{ballenegger-etal} or \cite[Lemma
2]{macris-martin-pule}). Formula \eqref{electro-force-almost-simplified}
eventually reduces to \eqref{point-electro-force} since $\int_0^{p_1}\!\!\d
s_1 \int_0^{p_2}\!\!\d s_2 \delta(\widetilde{s_1}-\widetilde{s_2}) = p_1\,
p_2$.

The properties \eqref{density-invariance} and
\eqref{correlation-invariance} both follow from the fact that the loop's
self-energy in the activity \eqref{loop-activity} is invariant under a
shift of origin and the loop pairwise interaction $V(\L_i,\L_j)$ \eqref{V}
is invariant when the loops $\L_i$ and $\L_j$ have their origin shifted to
$\L_i^{[u_1]}$ and $\L_j^{[u_2]}$ with $u_1-u_2\in\mathbb{Z}$. The
restriction $u_1-u_2 \in \mathbb{Z}$ is the manifestation of the
Feynman--Kac equal-time constraint in $\Vc$ and of the quantum nature of
the photon field in $\Wm$, occurring through the function $\mathcal{Q}(\nK,
\widetilde{s_1}-\widetilde{s_2})$ \eqref{Q}: $\Wm$ is unchanged because
$\mathcal{Q}\big(\nK, (\widetilde{s_1+u_1})-(\widetilde{s_2+u_2})\big) =
\mathcal{Q}(\nK, \widetilde{s_1}-\widetilde{s_2})$ when
$u_1-u_2\in\mathbb{Z}$ by periodicity of the function $s\mapsto
\mathcal{Q}(\nK, s)$.

\section{Screening of the resummed interaction $\Phi$}\label{appx:Phi-screened}
The classical Debye--H\"uckel potential $\Phi^\text{class}(x_1,x_2,\y)$ for
slab geometry has been extensively studied in
\cite{buenzli-martin-classical}. Because of the wall constraint on the
screening clouds, this potential does not decay exponentially fast as would
be the case in the bulk, but still has an integrable tail $\sim y^{-3}$
along the wall.\footnote{This was noticed long ago in
Ref. \cite{jancovici-wall}, see also
\cite[Sec. III.C.2]{martin-sum-rules}.} This implies that its transverse
Fourier transform $\Phi^\text{class}(x_1,x_2,\k)$ is finite at $\k=\0$, see
Formula (A.11) of \cite{buenzli-martin-classical}. In the sequel, we infer
that the same property remains true for the screened potential between
loops $\Phi(1,2,\k)$ defined by \eqref{Phiel-integral-def}. The only
difference between the Coulomb potential $v(\r_1-\r_2)$ for point charges
and $\Vel(\b 1,\b 2)$ is the extension of the loops that generates
additional multipole interactions. To disentangle the monopole interaction
from the multipole contributions, we proceed with the same method as in
Sec. V.B.2 of \cite{brydges-martin} and only sketch the
arguments. Introducing the multipole operator
\begin{align}
	\mathcal{M}_i = \int_0^{p_i}\!\!\!\d s\ \sum_{l=1}^\infty
	\frac{\big[\lambda_{\gamma_i} \X_i(s) \cdot \nabla_{\r_i}\big]^l}{l!},
	\quad i=1,2,
\end{align}
the loop interaction is decomposed into its charge--charge (cc),
charge--multipole (cm, mc) or multipole--multipole (mm) components: $\Vel =
\Vel_\text{cc} + \Vel_\text{cm} + \Vel_\text{mc} + \Vel_\text{mm}$, where
\begin{alignat}{2}
	&\Vel_\text{cc}(\b 1, \b 2) = p_1 p_2\ v(\r_1-\r_2), &\quad
	&\Vel_\text{cm}(\b 1, \b 2) = p_1\,\mathcal{M}_2\ v(\r_1-\r_2), \notag
	\\&\Vel_\text{mc}(\b 1, \b 2) = \mathcal{M}_1\, p_2\ v(\r_1-\r_2),
	&\quad &\Vel_\text{mm}(\b 1, \b 2) = \mathcal{M}_1\, \mathcal{M}_2\
	v(\r_1-\r_2).
\end{alignat}
Summing the chains of $\Vel$ to form $\Phi$ amounts to summing all possible
chains with bonds $\Vel_\text{cc}$, $\Vel_\text{cm}$, $\Vel_\text{mc}$, and
$\Vel_\text{mm}$. Summing first the pure $\Vel_\text{cc}$ chains builds the
classical Debye--H\"uckel potential $\Phi^\text{class}$ with screening
length $\kappa^{-1}(x) = [4\pi\beta \sum_\gamma\sum_p \int \D(\X) p^2
e_{\gamma}^2 \rho(x,\chi)]^{-1/2}$.\footnote{This screening length reduces
to the classical expression $[4\pi\beta \sum_{\gamma} e_\gamma^2
\rho(x,\gamma)]^{-1/2}$  when exchange effects
are disregarded \cite[Sec. V.B.2]{brydges-martin}.} One is
then left with the screened bonds
\begin{alignat}{2}
	&F_\text{cc} = -\beta e_{\gamma_1} e_{\gamma_2}\, p_1 p_2\
	\Phi^\text{class}, &\ &F_\text{cm} = -\beta e_{\gamma_1} e_{\gamma_2}\
	p_1\, \mathcal{M}_2\ \Phi^\text{class}, \notag
	\\&F_\text{mc} = -\beta e_{\gamma_1} e_{\gamma_2}\, \mathcal{M}_1\,
	p_2\ \Phi^\text{class}, &\ &F_\text{mm} = -\beta e_{\gamma_1}
	e_{\gamma_2}\, \mathcal{M}_1\, \mathcal{M}_2\ \Phi^\text{class}.
\end{alignat}
Finally, $\Phi$ is built from convolution chains of these screened bonds
subject to excluded convolution rules with respect to $F_\text{cc}$. One
sees that
\begin{align}
	F_\text{cm}(\b 1, \b 2) = -\beta e_{\gamma_1} e_{\gamma_2} p_1
	\!\!\!\int_0^{p_2}\!\!\! \d s_2 \left[ \Phi^\text{class}(\r_1,
	\r_2+\lambda_{\gamma_2} \X_2(s_2)) - \Phi^\text{class}(\r_1,\r_2)
	\right]
\end{align}
is integrable in the $\y$ direction and so has a finite transverse Fourier
transform at $\k=\0$. The same holds for the other screened bonds and their
chain convolutions, hence the result \eqref{phi-decay-prop}. These
considerations apply to the various types of screened potentials considered
in this work, \eg, the screened potential of the single plate systems
$\Phi_{A}^0,\;\Phi_{B}^0$ and the potentials
$\widetilde\Phi_{AA},\;\widetilde\Phi_{BB}$ occurring in Subsection 6.1.

\section{Dipole potential $\W_{AB}$ at order $\Order(d^{-1})$}\label{appx:Wm-asymptotic}
From \eqref{WAB-Fourier3D} and \eqref{Wm-class-asymptotic}, the asymptotic
form of $\W_{AB}(1,2,\frac{\q}{d})$ is
\begin{align}
	&\W_{AB}(1,2,\frac{\q}{d}) \stackrel{d\to\infty}{\sim} \frac{1}{d}
	\int\!\! \frac{\d q_1}{2\pi}\e^{-i q_1 x}
	\W(\chi_1,\chi_2,\tfrac{q_1}{d}, \tfrac{\q}{d})\Big\vert_{x=1} \notag
	\\ &=\frac{1}{d} \frac{\lambda_{\gamma_1}\lambda_{\gamma_2}}{\beta
	\sqrt{m_{\gamma_1}m_{\gamma_2}} c^2} \int_0^{p_1}\!\!\!\d X_1^\mu(s_1)
	\!\!  \int_0^{p_2}\!\!\!\d X_2^\mu(s_2)\int\!\! \frac{\d
	q_1}{2\pi}\e^{-i q_1 x} \notag
	\\& \phantom{=} \big[q_1 X_1(s_1) + \q \cdot \Y_1(s_1)\big] \big[q_1
	X_2(s_2) + \q \cdot \Y_2(s_2)\big] \frac{4\pi}{q_1^2+ q^2}
	\delta_{\mu\nu}^\text{tr}(q_1, \q)\Big\vert_{x=1} \notag
	\\ &=\frac{1}{d} \frac{\lambda_{\gamma_1}\lambda_{\gamma_2}}{\beta
	\sqrt{m_{\gamma_1}m_{\gamma_2}} c^2} \int_0^{p_1}\!\!\!\d X_1^\mu(s_1)
	\!\!  \int_0^{p_2}\!\!\!\d X_2^\mu(s_2) \notag
	\\& \phantom{=} \big[i X_1(s_1) \tfrac{\p}{\p {x}} + \q \cdot
	\Y_1(s_1)\big] \big[i X_2(s_2) \tfrac{\p}{\p {x}} + \q \cdot
	\Y_2(s_2)\big] v^{\mu\nu}(x, \q)\Big\vert_{x=1}
	\label{WAB-asymptotic-final}
\end{align}
where
\begin{align}
	v^{\mu\nu}(x,\q) &= \int\!\!\frac{\d q_1}{2\pi} \e^{i q_1 x}
	\frac{4\pi}{q_1^2+ q^2} \delta_{\mu\nu}^\text{tr}(q_1, \q) \notag
	\\&= \frac{\pi}{q}\e^{-q|x|} \times
        \begin{cases}
            \delta_{\mu\nu} + q|x|, &\nu=\mu=1,
            \\ \delta_{\mu\nu}-iq_\mu x, &\mu\neq 1, \nu=1,
            \\ 2\delta_{\mu\nu}-(1+q|x|)\frac{q_\mu q_\nu}{q^2}, &\mu\neq
            1, \nu\neq 1
        \end{cases} \label{vT-x-k}
\end{align}
is the partial Fourier transform of the transverse Coulomb potential. The
final result (Formula (5.88) in \cite{buenzli-thesis}) is obtained by
working out the derivatives in \eqref{WAB-asymptotic-final}.

\end{document}